\documentclass[11pt]{prop_axis} 

\usepackage[super,comma]{natbib} 
\usepackage{apjfonts} 
\usepackage{graphicx} 
\usepackage{amssymb,amsmath} 
\usepackage{color} 
\usepackage{enumitem} 
\usepackage{wrapfig} 
\usepackage{tikz} 

\usepackage{multirow} 
\usepackage{tabularx} 
\usepackage{array} 

\usepackage{hyperref} 
\hypersetup{
    colorlinks,
    linkcolor={blue!80!black},
    citecolor={blue!80!black},
    urlcolor={blue!80!black}
}

\usepackage{sidecap}
\sidecaptionvpos{figure}{c} 

\newcolumntype{L}[1]{>{\raggedright\let\newline\\\arraybackslash\hspace{0pt}}p{#1}}
\newcolumntype{C}[1]{>{\centering\let\newline\\\arraybackslash\hspace{0pt}}p{#1}}
\newcolumntype{R}[1]{>{\raggedleft\let\newline\\\arraybackslash\hspace{0pt}}p{#1}}
\renewcommand\arraystretch{1.3}

\definecolor{headcolor}{rgb}{0.65,0.65,0.65}
\usepackage{fancyhdr}
\renewcommand{\topmargin}{-9mm}

\renewcommand{\textfloatsep}{4mm}

\newcommand{\lem}{\textit{LEM}}

\newcommand{\chandra}{\textit{Chandra}}

\newcommand{\xmm}{\textit{XMM-Newton}}
\newcommand{\XMM}{\textit{XMM-Newton}}

\newcommand{\Suzaku}{\textit{Suzaku}}

\newcommand{\athena}{\textit{Athena}}

\newcommand{\xrism}{\textit{XRISM}}

\newcommand{\JWST}{\textit{JWST}}

\newcommand{\bsf}{\sffamily\bfseries}

\definecolor{callout}{rgb}{0.25,0.40,0.85}
\definecolor{synergies}{rgb}{0.20,0.45,0.99}
\definecolor{methods}{rgb}{0.20,0.70,0.45}

\definecolor{calllem}{rgb}{0.20,0.45,0.99}
\definecolor{tabledef}{rgb}{0.95,0.95,0.95}
\definecolor{tablealt}{rgb}{0.77,0.80,1.0}
\definecolor{tablelem}{rgb}{0.80,0.85,1.0}
\definecolor{whitelem}{rgb}{1.0,1.0,1.0}
\definecolor{greenlem}{rgb}{0.7,1.0,0.7}

\begin{document}

\baselineskip=13.2pt
\sloppy
\pagenumbering{roman}
\thispagestyle{empty}


\title{\textcolor{headcolor}{\LARGE\bsf 
Exploring chemical enrichment of the intracluster medium with the Line Emission Mapper}}
\maketitle
\vspace*{-10mm}

\begin{tikzpicture}[remember picture,overlay]
\node[anchor=north west,yshift=2pt,xshift=2pt]%
    at (current page.north west)
    {\includegraphics[height=20mm]{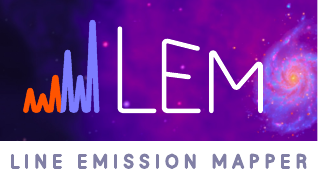}};
\end{tikzpicture}

\vspace*{-13mm}
\begin{center}
\begin{minipage}{17.5cm}
\hspace{-5mm}
\centering
Fran\c{c}ois~Mernier$^{1,2}$,
Yuanyuan~Su$^{3}$,
Maxim~Markevitch$^{1}$,
Congyao~Zhang$^{4}$,
Aurora~Simionescu$^{5,6,7}$,
Elena~Rasia$^{8,9}$,
Sheng-Chieh~Lin$^{3}$,
Irina~Zhuravleva$^{4}$,
Arnab~Sarkar$^{10}$,
Ralph~P.~Kraft$^{11}$,
Anna~Ogorzalek$^{1,2}$,
Mohammadreza~Ayromlou$^{13}$,
William~R.~Forman$^{11}$,
Christine~Jones$^{11}$,
Joel~N.~Bregman$^{14}$,
Stefano~Ettori$^{15,16}$,
Klaus~Dolag$^{17,12}$,
Veronica~Biffi$^{9}$,
Eugene~Churazov$^{18,19}$,
Ming~Sun$^{20}$,
John~ZuHone$^{11}$,
\'{A}kos~Bogd\'{a}n$^{11}$,
Ildar~I.~Khabibullin$^{17,18}$,
Norbert~Werner$^{21}$,
Nhut~Truong$^{1,22}$,
Annalisa~Pillepich$^{23}$,
Priyanka~Chakraborty$^{11}$, 
Stephen~A.~Walker$^{20}$, 
Mark~Vogelsberger$^{10}$,
Mohammad~S.~Mirakhor$^{20}$

\end{minipage}
\end{center}

\vfill



{\footnotesize
\noindent
$^{1}$~~NASA Goddard Space Flight Center, Code 662, Greenbelt, MD 20771, USA\\
$^{2}$~~Department of Astronomy, University of Maryland, College Park, MD 20742-2421, USA\\
$^{3}$~~Department of Physics and Astronomy, University of Kentucky, 505 Rose Street, Lexington, KY 40506, USA\\
$^{4}$~~Department of Astronomy and Astrophysics, The University of Chicago, Chicago, IL 60637, USA\\
$^{5}$~~SRON Netherlands Institute for Space Research, Niels Bohrweg 4, NL-2333 CA Leiden, the Netherlands\\
$^{6}$~~Leiden Observatory, Leiden University, PO Box 9513, NL-2300 RA Leiden, the Netherlands\\
$^{7}$~~Kavli Institute for the Physics and Mathematics of the Universe, University of Tokyo, Kashiwa 277-8583, Japan\\
$^{8}$~~IFPU - Institute for Fundamental Physics of the Universe, Via Beirut 2, I-34014 Trieste, Italy\\
$^{9}$~~INAF Osservatorio Astronomico di Trieste, via Tiepolo 11, I-34131, Trieste, Italy\\
$^{10}$~Kavli Institute for Astrophysics and Space Research, Massachusetts Institute of Technology, 77 Massachusetts Ave, Cambridge, MA 02139, USA\\
$^{11}$~Center for Astrophysics $|$ Harvard \& Smithsonian, 60 Garden St, Cambridge, MA 02138, USA\\
$^{12}$~Max Planck Institute for Astrophysics, Karl-Schwarzschild-Str. 1, Garching bei M\"{u}nchen 85741, Germany\\
$^{13}$~Universit\"{a}t Heidelberg, Zentrum f\"{u}r Astronomie, Institut f\"{u}r theoretische Astrophysik, Albert-Ueberle-Str. 2, Heidelberg 69120, Germany\\
$^{14}$~Department of Astronomy, University of Michigan, 311 West Hall, 1085 S. University Ave, Ann Arbor, MI, 48109-1107, USA\\
$^{15}$~INAF, Osservatorio di Astrofisica e Scienza dello Spazio di Bologna, via Piero Gobetti 93/3, I-40129 Bologna, Italy\\
$^{16}$~INFN, Sezione di Bologna, viale Berti Pichat 6/2, I-40127 Bologna, Italy\\
$^{17}$~Universit\"{a}ts-Sternwarte, Fakult\"{a}t f\"{u}r Physik, Ludwig-Maximilians-Universit\"{a}t M\"{u}nchen, Scheinerstr 1, D-81679 M\"{u}nchen, Germany\\
$^{18}$~Space Research Institute (IKI), Profsoyuznaya 84/32, 117997 Moscow, Russia\\
$^{19}$~Max Planck Institute for Astrophysics, Karl-Schwarzschild-Str. 1, D-85741 Garching, Germany\\
$^{20}$~Department of Physics and Astronomy, University of Alabama in Huntsville, Huntsville, AL 35899, USA\\
$^{21}$~Department of Theoretical Physics and Astrophysics, Faculty of Science, Masaryk University, Kotl\'{a}\v{r}sk\'{a} 2, CZ-611 37 Brno, Czech Republic\\
$^{22}$Center for Space Science and Technology, University of Maryland, Baltimore County, 1000 Hilltop Circle, Baltimore, MD 21250, USA\\
$^{23}$~Max-Planck-Institut f\"{u}r Astronomie, K\"{o}nigstuhl 17, D-69117 Heidelberg, Germany\\
\\
\phantom{${^52}$}~\textcolor{blue}{\bsf \href{https://lem-observatory.org}{lem-observatory.org}}\\
\phantom{${^52}$}~\textcolor{blue}{\bsf X / twitter: \href{https://www.twitter.com/LEMXray}{LEMXray}}\\
\phantom{${^52}$}~\textcolor{blue}{\bsf facebook: \href{https://www.facebook.com/LEMXrayProbe}{LEMXrayProbe}}\\
}

\centerline{\em White Paper, October 2023}

\clearpage
\twocolumn


\setcounter{page}{1}
\pagenumbering{arabic}

\section*{SUMMARY}
\label{sec:summary}

Synthesized in the cores of stars and supernovae, most metals disperse over cosmic scales and are ultimately deposited well outside the gravitational potential of their host galaxies. Since their presence is well visible through their X-ray emission lines in the hot gas pervading galaxy clusters, measuring metal abundances in the intracluster medium (ICM) offers us a unique view of chemical enrichment of the Universe as a whole. Despite extraordinary progress in the field thanks to four decades of X-ray spectroscopy using CCD (and gratings) instruments, understanding the precise stellar origins of the bulk of metals, and when the latter were mixed on Mpc scales, requires an X-ray mission capable of spatial, non-dispersive high resolution spectroscopy covering at least the soft X-ray band over a large field of view. In this White Paper, we demonstrate how the \textit{Line Emission Mapper} (\lem) probe mission concept will revolutionize our current picture of the ICM enrichment. Specifically, we show that \lem\ will be able to (i) spatially map the distribution of ten key chemical elements out to the virial radius of a nearby relaxed cluster and (ii) measure metal abundances in serendipitously discovered high-redshift protoclusters. Altogether, these key observables will allow us to constrain the chemical history of the largest gravitationally bound structures of the Universe. They will also solve key questions such as the universality of the initial mass function (IMF) and the initial metallicity of the stellar populations producing these metals, as well as the relative contribution of asymptotic giant branch (AGB) stars, core-collapse, and Type Ia supernovae to enrich the cosmic web over Mpc scales. Concrete observing strategies are also briefly discussed.

\section{Introduction}
\label{sec:intro}

The properties of clusters of galaxies throughout cosmic history are sensitive tracers of the underlying cosmology, structure formation, and hierarchical growth. Understanding how these largest gravitationally bound structures formed, grew, and further evolve is thus of major interest for astrophysics and cosmology as a whole.

%
%

\begin{figure*}[h]
\centering
\includegraphics[width=0.99\textwidth, trim={0cm 0cm 0cm 0cm},clip]{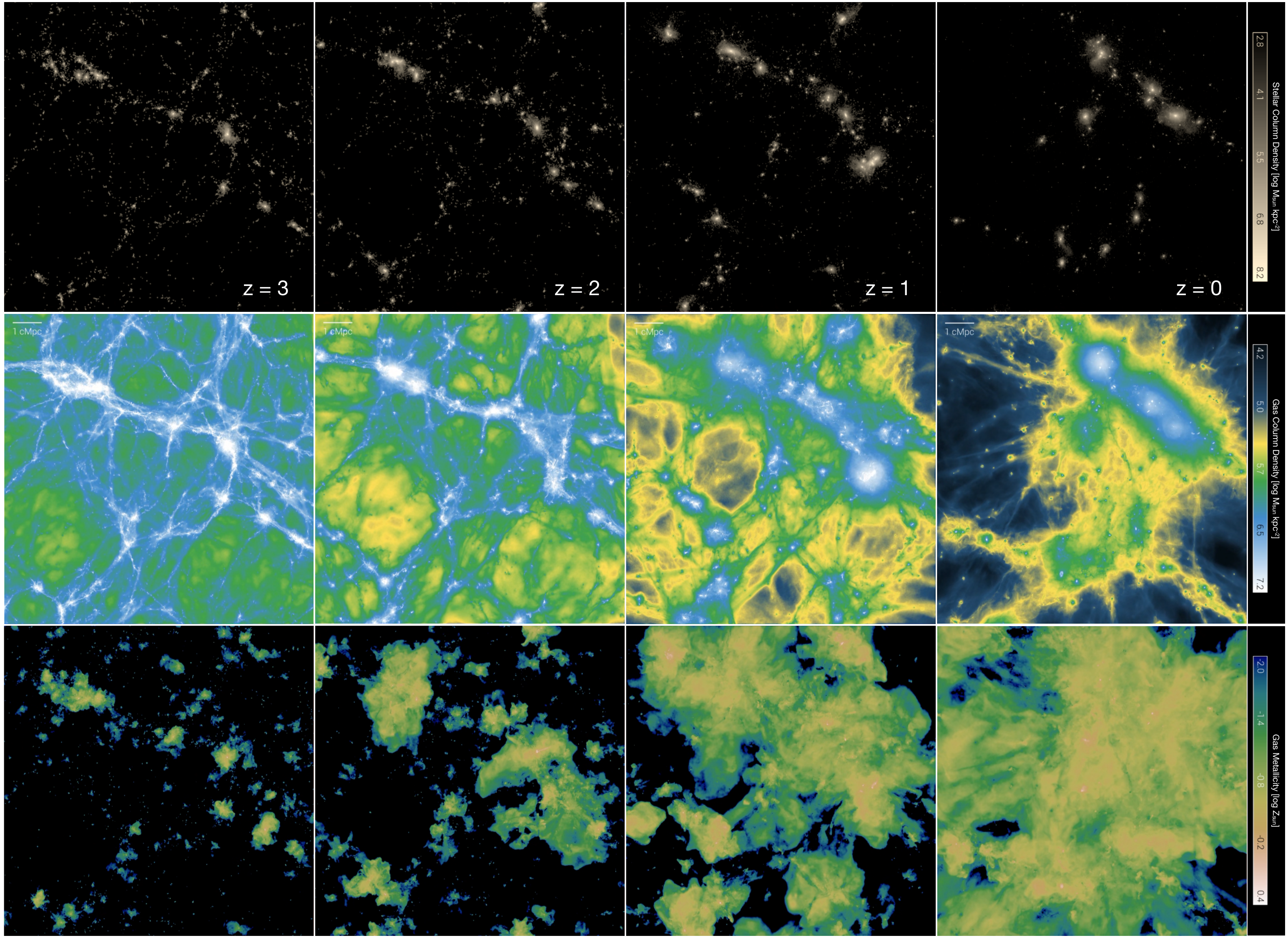}

\caption{Evolving structure of a 10~Mpc region as simulated by TNG100-1 of the IllustrisTNG project\cite{Nelson18,Pillepich18,Springel18,Naiman18,Marinacci18} from $z=5$ until now. At such large scales, the stellar content (\textit{top panel}) accounts for only a small fraction of the baryons. The hot-gas phase and its metal content (\textit{middle}: gas density; \textit{bottom}: gas metallicity), accessible essentially through X-ray (continuum and line) emission, are key for obtaining a full picture of the cycle of baryons and metals in the Universe.}

\label{fig:mill}
\end{figure*}

\begin{figure}[h]
\centering
\includegraphics[width=0.49\textwidth, trim={0cm 0cm 0cm 0cm},clip]{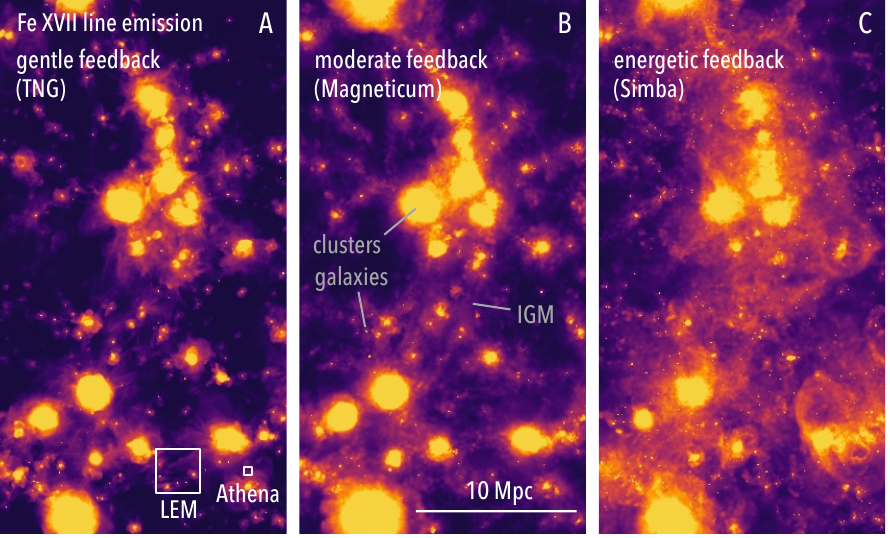}

\caption{Fe XVII line emission map simulated for cosmic large-scale structures with gentle feedback (\textit{left}, TNG), moderate feedback (\textit{middle}, Magneticum), and energetic feedback (\textit{right}, Simba) from the CAMELS project\cite{Villaescusa21}. With its large grasp and high energy resolution capabilities, \lem\ is the perfect instrument to probe the hot gas in (and even between) clusters\cite{Markevitch23}.}

\label{fig:filaments}
\end{figure}

Revealed thanks to their emission lines in X-ray spectra of groups and clusters of galaxies, the presence of metals in the hot intracluster medium (ICM) is a key piece of the puzzle in our understanding of the chemical evolution of structures at Mpc scales. While metals inside galaxies are at the origin of rocky planets and even life, finding the majority of these chemical elements outside galaxies poses profound questions on the cosmic epoch of the enrichment of our Universe at large scales, as well as on the properties of their parent stellar populations\cite{Biffi18,Mernier18c,Mernier22b}. Comprising $\sim$80\% of the total baryonic mass in clusters, the ICM phase represents, in fact, the most valuable territory to explore and understand the cycle of metals as a whole (Fig.~\ref{fig:mill})\footnote{\href{https://www.tng-project.org/media/}{https://www.tng-project.org/media/}}. Inside the virial radius of massive clusters, this hot gas constitutes a remarkable fossil record of the enrichment of the Universe. Due to the high temperature of the ICM ($10^7$--$10^8$~K), metals therein (and their associated chemical abundances) are accessible only in the X-ray band. Even more spectacularly, the relatively simple physical conditions of this plasma (i.e. in collisional ionization equilibrium, optically thin at almost all X-ray energies) means that the ICM is the astrophysical system where chemical abundances can be, by far, the most accurately constrained---even better than in our own Solar System\cite{Mernier18b,Simionescu19}.

Four decades of dedicated observations with past and current X-ray telescopes\cite{Werner08,Mernier18c}, however, taught us that this full potential can be reached only via a wide-field mission capable of high-resolution X-ray spectral mapping. Currently, reliable abundance measurements at CCD resolution ($\sim$120~eV) are in fact limited to the cases of well known and/or isolated dominant spectral lines (e.g. Fe-K at $\sim$6.7~keV, Si-K and at $\sim$2~keV) in spectra of excellent statistics and limited background contamination. Coupled with the uncertain contamination of the X-ray foreground (coming primarily from the diffuse emission of our own Milky Way), the extraordinary spectral complexity of the Fe-L line forest, at soft X-ray energies, renders the analysis of fainter and/or cooler regions of the ICM extremely challenging. Even in bright cluster cores, accurate abundance measurements are limited to a couple of elements, providing us only with a partial connection between large scale structures and the chemical feedback from their stellar populations. The recent launch of \xrism, and particularly its key instrument Resolve (with a spectral resolution of 5--7~eV), will arguably pave the way towards the era of non-dispersive high resolution X-ray spectroscopy\cite{XRISM20}. Its large point spread function and moderate effective area, however, inevitably introduces limitations that will restrict our ability to probe the spatial distribution and extent of metals in the ICM. Measurements in cluster outskirts, in particular, will be very challenging to perform accurately, meaning that most of the cluster volume and mass will remain essentially unexplored for the years to come. Beyond \xrism, the extraordinary spectral resolution (3--4~eV) and effective area of the X-IFU instrument onboard \textit{NewAthena}\footnote{The \textit{NewAthena} specifications mentioned here might slightly change by the mission adoption. They are to the best of our knowledge at the time this paper is written. For clarity, references relevant to the mission before its redefinition in late 2022 will be mentioned as \athena.} will enable much more detailed mapping of the metal abundances in nearby cluster cores, and transform our understanding of the global properties of high-redshift systems\cite{Barret23}; however, its limited field of view (FoV; $\lesssim$5$'$ diameter) will make full coverage of any nearby system very time-consuming. An ideal observatory for the above science would combine the high spectral resolution and excellent spatial resolution of \textit{NewAthena} with the large FoV of \XMM.

Proposed as a mission concept for the NASA 2023 Astrophysics Probes call, \lem\ is an extremely promising space observatory\cite{Kraft22}. It consists of a micro-calorimeter array spanning a large FoV of $30 \times 30$~arcmin and capable of reaching an energy resolution of 1~eV (central $7\times 7$~arcmin) to 2~eV (rest of the FoV) in the 0.2--2~keV band. The X-ray mirrors of the telescope will provide excellent collecting area ($\sim$1600~cm$^2$) and angular resolution ($\sim$15$''$), of the order of the pn instrument onboard \XMM. While the primary science topic to be addressed by \lem\ is the circumgalactic medium and its physical properties\cite{Nelson23,Truong23,Schellenberger23,ZuHone23,Bogdan23}, this mission is, in fact, ideally suited for detailed studies of the ICM across the entire volume of any nearby system. First, \lem\ will be able to characterize with unprecedented accuracy the velocity fields in cluster outskirts at small scales (turbulence) and large scales (bulk motions) simultaneously, proving crucial insights on the assembly of groups and clusters\cite{Zhang24}. Second, the large grasp of \lem, coupled with its microcalorimeter resolution, makes it a key mission to probe the gas distribution in and even outside clusters, hence allowing for a better understanding of the feedback history of large scale structures\cite{Markevitch23} (Fig.~\ref{fig:filaments}).

As pointed out by the Astro2020 Decadal Survey\footnote{https://nap.nationalacademies.org/catalog/26141/pathways-to-discovery-in-astronomy-and-astrophysics-for-the-2020s}, the concept of enrichment is ubiquitous in our quest to understand Cosmic Ecosystems (1.1.3), Stellar and Black Hole Feedback (2.3.2), Multi-Scale Cosmic Flows of Gas (2.3.3), as well as the deep connection between the three. Operating at the X-ray regime---and thus remarkably complementary to enrichment studies of stellar populations and cold gas phases in galaxies performed by \JWST\ and future NASA Great Observatories (e.g. \textit{LUVOIR}), \lem\ will provide the missing piece necessary to solve the puzzle of how gas and metals flow into, through, and out of galaxies (Appendix D-Q2 of the Decadal Survey). 

In this White Paper, we demonstrate how the \lem\ imaging microcalorimeter mission will push our understanding of the chemical history of the ICM to the next level. This work is divided as follows. Section~\ref{sec:questions} provides an overview and introduces the still open questions on chemical enrichment. Section~\ref{sec:LEM} places \lem\ into this context and reports observational predictions obtained via three various approaches. Section~\ref{sec:conclusion} briefly summarizes the \lem\ prospects and addresses observing strategies on chemical studies of the ICM.

\begin{figure*}[t]
\centering
\includegraphics[width=0.447\textwidth, trim={1.5cm 0.3cm 1.9cm 0.3cm},clip]{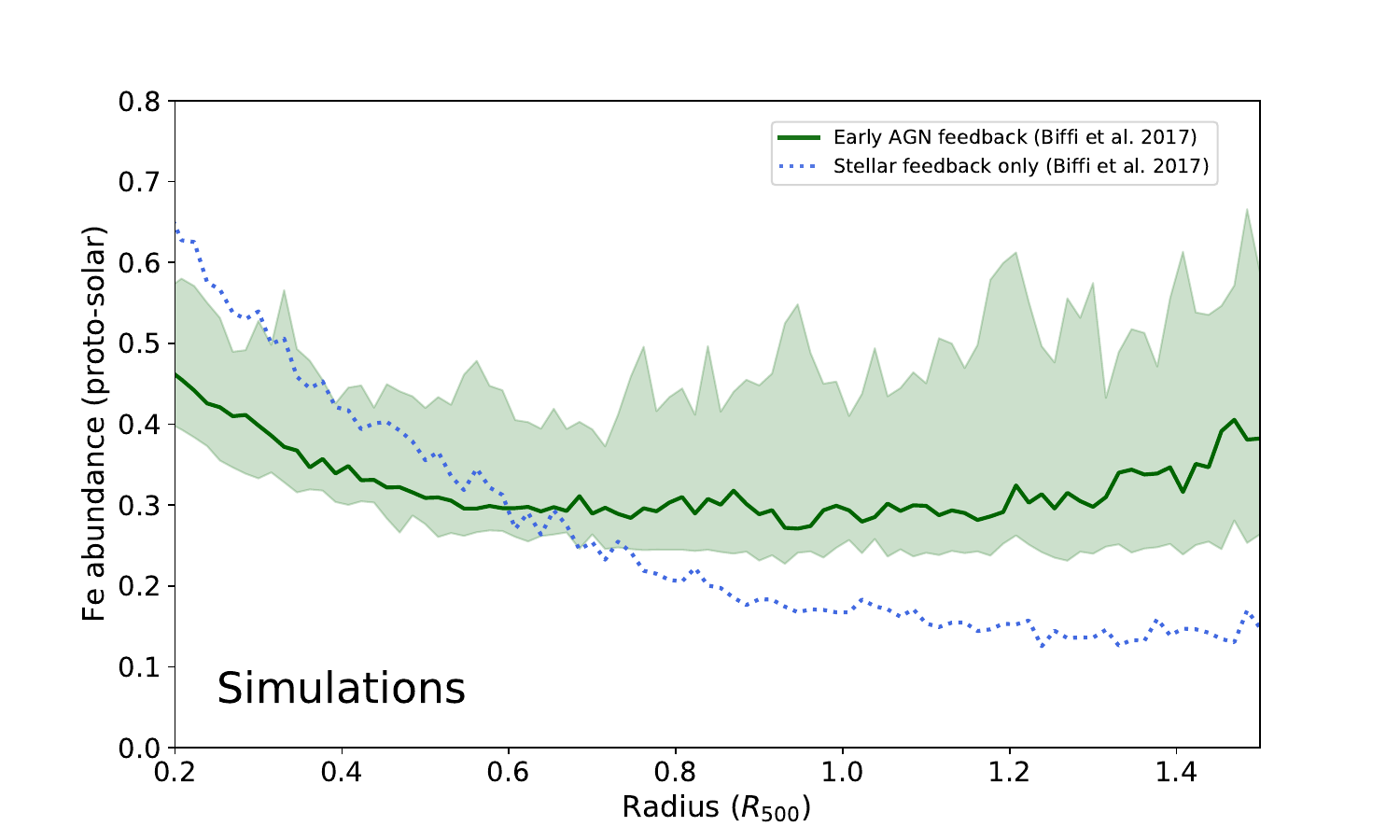}
\includegraphics[width=0.54\textwidth, trim={0cm 0cm 0cm 0cm},clip]{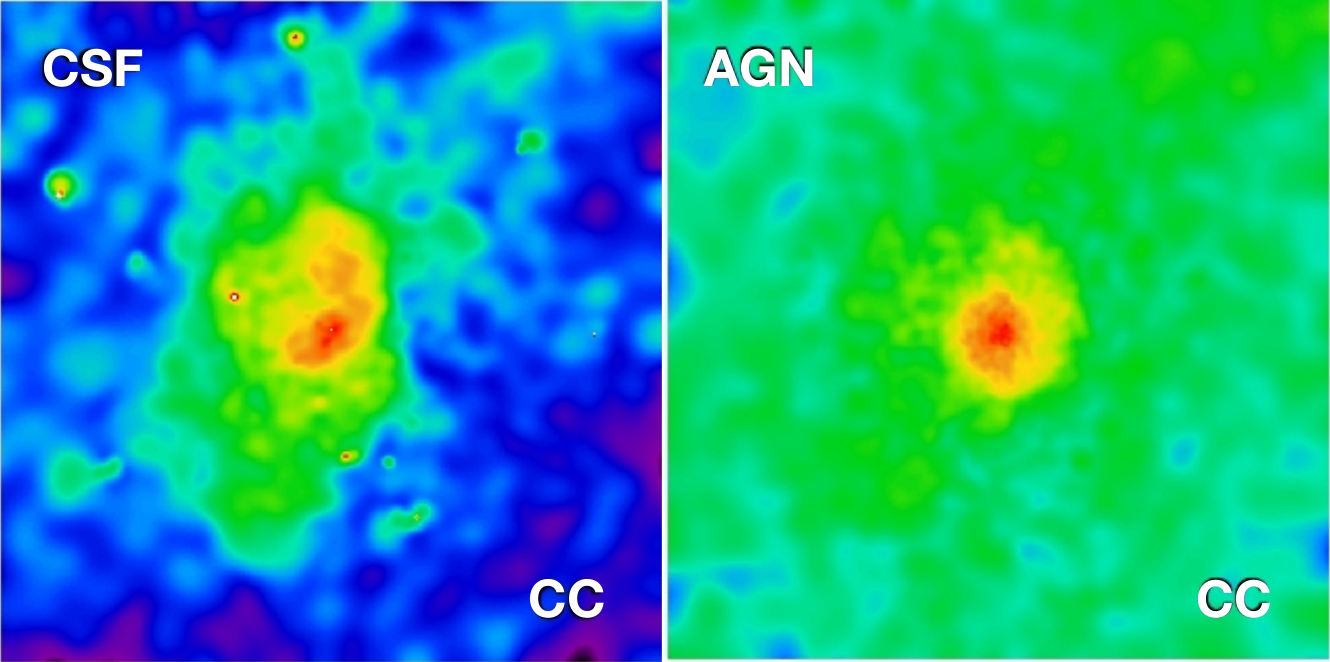}

\caption{Emission-weighted Fe abundance spatial distribution as predicted by the Dianoga cosmological simulation, comparing two feedback prescriptions: star formation only (CSF) and the addition of early feeback from supermassive black holes (AGN). \textit{Left:} Abundance radial profile (adapted from Biffi et al. 2018)\cite{Biffi18}. \textit{Right:} Projected abundance map (reprinted with permission from Biffi et al. 2017)\cite{Biffi17}. A SMBH dominant enrichment mechanism occurring early on mixes metals more uniformly, whereas stellar feedback only is expected to result in a more patchy, concentrated metallicity distribution\cite{Biffi17}.}

\label{fig:simulations}
\end{figure*}

\section{Chemical enrichment of the ICM: state of the art and open questions}\label{sec:questions}

Over the last decades, X-ray space observatories and state-of-the-art hydrodynamical cosmological simulations enabled enormous progress in understanding the journey of metals from their creation in stars and supernovae to their mixing at Mpc scales\cite{Werner08,Biffi18,Mernier18c}. Tracing back the history of the enrichment and the stellar population(s) at its origin, however, remain key challenges for the decades to come. We discuss these two aspects of major astrophysical interest further below.

\subsection{When did galaxies enrich the ICM?}\label{subsec:when}

Crucial clues regarding the history of the ICM enrichment hide in the outskirts of galaxy clusters. In regions close to the virial radius, the gas has not had time yet to fully accrete nor virialize with the central ICM. Measuring its metal content and chemical composition constitutes, thus, a formidable opportunity to determine whether the bulk of the enrichment occurred before or after clusters have assembled. At CCD spectral resolution, Fe is by far the easiest element to probe in the ICM (due to its high relative abundance and its prominent Fe-K line at 6.7~keV). Observations of a handful of massive clusters and low-mass galaxy groups point toward an isotropic floor of Fe abundance at $\sim$0.3 Solar\cite{Werner13,Urban17,Sarkar22} (in units of Lodders et al. 2009)\cite{Lodders09}. Type Ia supernovae (SNIa) are the primary source of Fe-peak elements, therefore the above results not only predict a strong enrichment level in the intergalactic medium, but also suggests that most SNIa products must have been synthesized then released outside galaxies \textit{before} the latter started to assemble into clusters. This is (at least qualitatively) in line with some of the state-of-the-art cosmological hydrodynamical simulations\cite{Biffi17,Biffi18a,Angelinelli22}, showing that a prescription of early supermassive black-hole (SMBH) feedback is necessary to stir and mix metals homogeneously through the ICM and ultimately form a flat Fe abundance profile in cluster outskirts (Fig.~\ref{fig:simulations} left). Estimated at $z > 2$--3, the epoch of this pre-enrichment scenario coincides with the cosmic peak of star formation\cite{Madau14} and of SMBH activity\cite{Hickox18}. This current paradigm, however, is confronted by four crucial questions. 

\subsubsection{How ``early'' is the ICM Fe enrichment?} 

A pure pre-enrichment scenario requires SMBH feedback processes that are able to disperse metals uniformly through the early (not yet virialized) intergalactic medium, hence making Fe abundance in cluster outskirts remarkably homogeneous at all spatial scales (Fig.~\ref{fig:simulations} right). While results from \Suzaku\ in the Perseus cluster outskirts suggest the Fe abundance floor to be homogeneous over >100--200 kpc scales, the difficulty of detecting emission lines at weak signal/background ratio with CCD spectral resolution implies that spatial metallicity fluctuations at smaller scales remain virtually unexplored. In fact, SNIa are seen to explode at higher rates in clusters than in the field\cite{Friedmann18} and a significant fraction of them could take up to a few Gyr to explode\cite{Maoz12}. This more recent contribution of Fe ejected from galaxies during or after cluster assembly (by recent stellar feedback directly or galaxy dynamics in the ICM) to the bulk of cluster enrichment remains widely unknown. 

\begin{figure*}[h]
\centering
\includegraphics[width=0.74\textwidth, trim={0cm 0cm 0cm 0cm},clip]{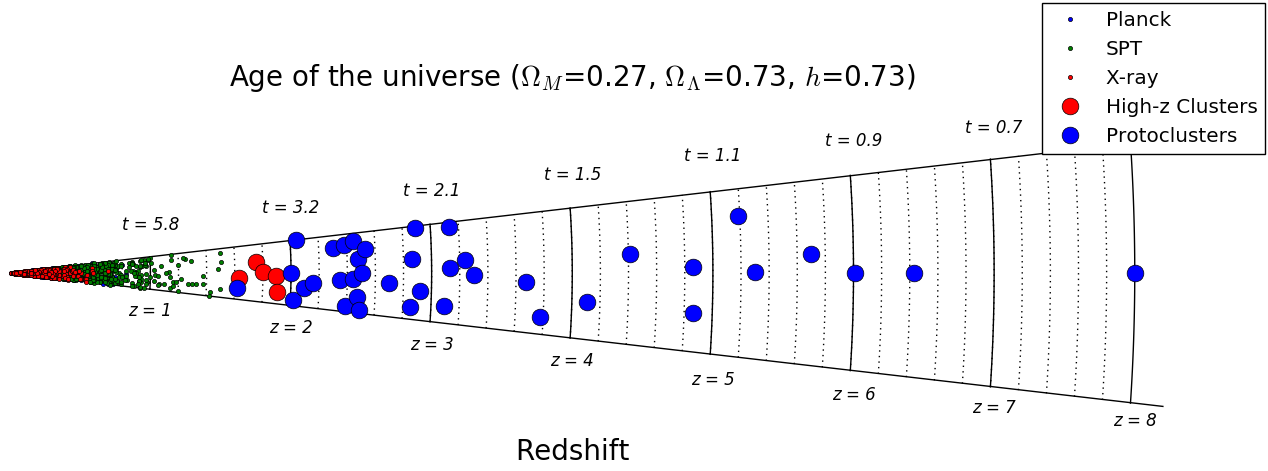}

\caption{The distribution of redshifts of clusters and protoclusters selected from the literature. Reprinted from Overzier (2016)\cite{Overzier16}.}

\label{fig:proto}
\end{figure*}

\subsubsection{How universal is the ICM Fe enrichment?}

The ultimate consequence of a pure pre-enrichment is that the gas of every nearby astrophysical system is expected to be considerably enriched with metals. In other words, pristine gas should not exist anymore in today's Universe: the Fe content in the outskirts of all galaxies, groups, and clusters should be universal and exhibit a common value of $\sim$0.3 Solar. This prediction corroborates results from cosmological simulations, which find a consistent level of enrichment between groups and clusters over their integrated volume\cite{Truong19,Angelinelli23}. Current CDD observations suggest similar results, however they remain limited to the bright central regions of relaxed clusters\cite{Mernier18a}. Whether groups are enriched at the same level as clusters through their entire volume (and by extension in their outskirts) is, in fact, not a trivial question due to the non-gravitational processes at play in these systems exhibiting shallower potential wells (e.g. higher stellar fraction and lower gas fraction due to the higher relative efficiency of SMBH feedback)\cite{Eckert21,Lovisari21}. Moreover, the extraordinary difficulty to constrain Fe abundances in cooler plasmas with CCD spectral resolution (due to the Fe-L forest being largely unresolved by current instruments) makes groups largely unexplored chemically\cite{Gastaldello21}.

\subsubsection{What is the ICM enrichment history of \ \ \ \ \ $\alpha$-elements?}

As stated earlier, Fe has been produced continuously by SNIa, exploding with a delay of a up to few Gyrs after star formation. On the other hand, $\alpha$-elements (particularly O, whose H-like emission line is remarkably bright in the X-ray band) are produced primarily by core-collapse supernovae (SNcc) which explode rapidly (a few tens of Myr at most) after the formation of their massive progenitors in star forming regions\cite{Nomoto13}. Whether SMBH feedback was fast enough to eject these lighter metals outside galaxies (hence preventing them to be locked back into stars) is an open and largely unexplored question. Only a handful of studies have been attempting to measure the radial distribution of the $\alpha$/Fe ratios at CCD resolution, whose diverging results keep us far from a clear consensus\cite{Simionescu15,Mernier17,Sarkar22}. Measuring for the content and spatial distribution of O at Mpc scales with unprecedented accuracy provides a unique opportunity to probe feedback timescales spanning the entire cosmic web at the scale of the Universe. Not only will it complete our view on the enrichment of the Universe, but it will also be a crucial test for future cosmological simulations.

\subsubsection{Can we trace back the enrichment history down to the era of protoclusters?}

A growing number of galaxy cluster progenitors, the so-called ``protoclusters", have been discovered at $z>2$ by galaxy-based surveys\cite{Steidel05,Matsuda05,LeFevre15,Toshikawa16} (Fig.~\ref{fig:proto}). Their detection significance, cluster masses, and dynamics are inferred mainly through the large-scale overdensity of galaxies, which can be highly uncertain and sensitive to projection effects and the variation of the field galaxy distribution. Our (still) small, heterogeneous and biased data set has already begun to refresh our view of galaxy evolution and the early stages of cluster evolution \cite{Hatch11, Casey15,Overzier16}. 
Compared to nearby mature clusters, protoclusters are more actively forming through accretion from the cosmic web\cite{Chiang13,Umehata19} and represent a key epoch of the build-up of the ICM, cluster metallicity, and cluster magnetic fields. In particular, the emergence of the ICM marks a critical stage in a (proto)cluster's virialization. The enrichment processes of galaxy clusters can be best probed by comparing the ICM composition in the nearby Universe with that at high redshift. If there is no evolution in the metallicity and abundance ratios of the ICM with redshift, it would strongly support the pre-enrichment scenario discussed above. If the level of ICM metallicity of nearby Universe exceeds that of the high redshift progenitors, it would manifest the role of ram pressure stripping of member galaxies, mass loss of their stellar components, etc., in enriching the ICM at a later epoch. One ultimate experiment is to compare the ICM composition of nearby clusters and protoclusters. Despite the wealth of information the ICM can provide, dedicated studies of protoclusters using direct X-ray emission and the Sunyaev-Zel'dovich (SZ) effect are almost non-existent. This is mainly due to the rarity of massive structures at high redshift, the large uncertainty in the timeline of ICM's emergence in simulations, and the required deep exposure time. Any ICM detection can be further contaminated by the presence of bright active galactic nuclei (AGN), star formation, and inverse Compton emission \cite{Wang16}, as protoclusters prevail in an epoch of galaxy fueling and are often discovered via radio galaxies as a tracer. To date, the only convincingly detected ICM in a protocluster is in the spiderweb field at $z=2.16$ \cite{Tozzi22,DiMascolo23}. A deep (700\,ks) \chandra\ observation has revealed that it contains a roughly symmetric diffuse emission within a radius of $\sim$100~kpc (11.6$^{\prime\prime}$) centered on the radio galaxy J1140-2629 (the Spiderweb galaxy; $3.5\times10^{13}$\,M$_{\odot}$). Its soft emission is consistent with thermal bremsstrahlung from a hot ICM with a temperature of $kT \simeq 2.0$\,keV, while its metallicity cannot be constrained. Although, on the other hand, the insensitivity of SZ to cosmological dimming appears as an advantage to detect ICM emission in protoclusters, that alternative method can hardly separate the signal of radio sources and of dust (which might be abundant in $z \sim 2$ objects). Moreover, the SZ signal cannot reveal the ICM metal content. Measurements of the latter can only be made through X-ray observations, but requires prohibitively long exposure times for current X-ray instruments. \\

To answer these four major questions, one needs an observatory that is capable of (i) exquisite spectral resolution to improve our abundance measurement accuracy (not only to resolve the Fe-L forest, but also to isolate the true ICM emission by separating these slightly redshifted lines from the rest-frame foreground Milky Way lines), (ii) small point-spread function to probe scales below 100 kpc in the ICM, and (iii) a wide FoV to capture the entire cluster volume at once. As further demonstrated in Sect.~\ref{sec:LEM}, \lem\ will take advantage of such unique capabilities to measure the spatial distribution of Fe, O, and the O/Fe ratio radially (from cluster cores to outskirts) and via 2D mapping to determine when and through which channels galaxies enriched the ICM. Such measurements will be critical towards constraining the complex physics of cosmological simulations, which still predict factors of ~2 different O/Fe ratios in cluster and group outskirts\cite{Vogelsberger18}.

\begin{figure*}[h]
\centering
\includegraphics[width=0.48\textwidth, trim={0.5cm 0cm 0cm 0cm},clip]{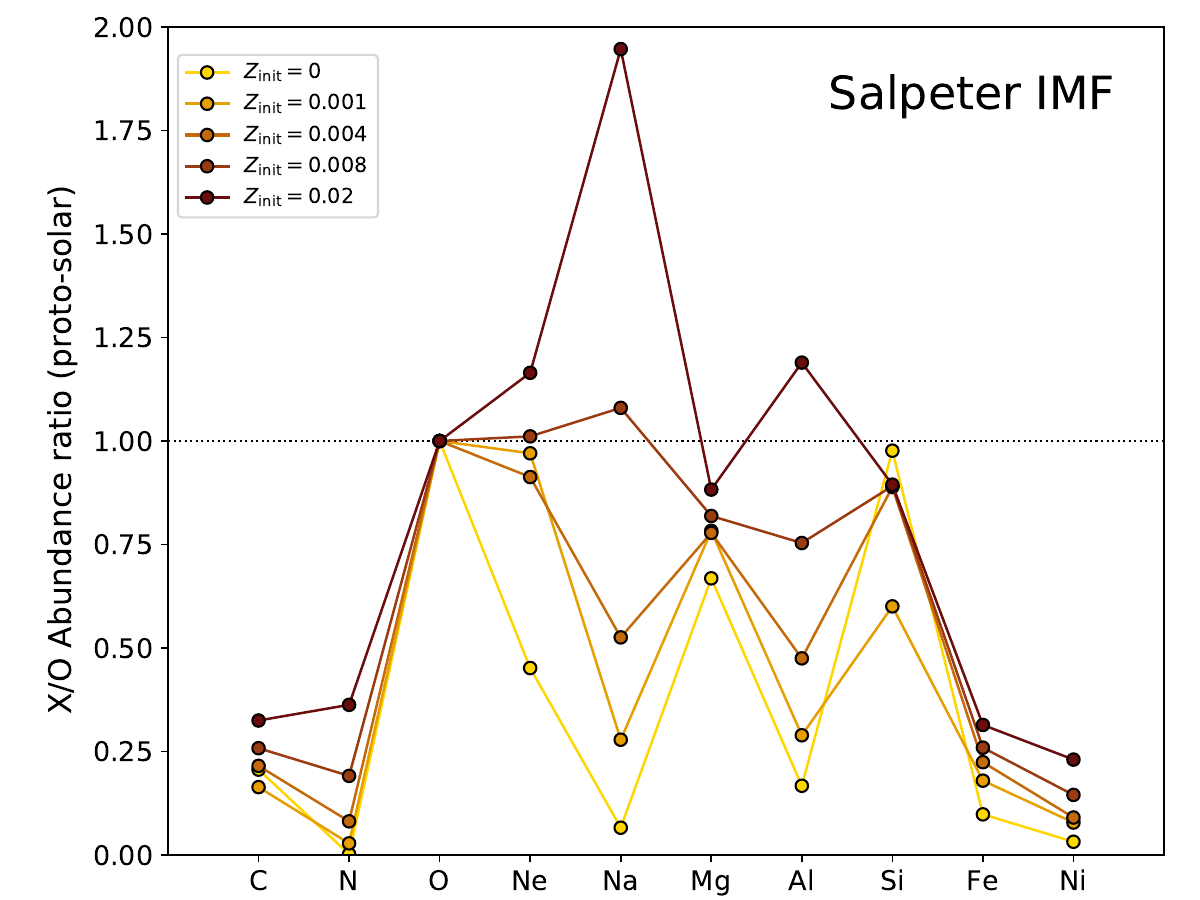}
\includegraphics[width=0.48\textwidth, trim={0.5cm 0cm 0cm 0cm},clip]{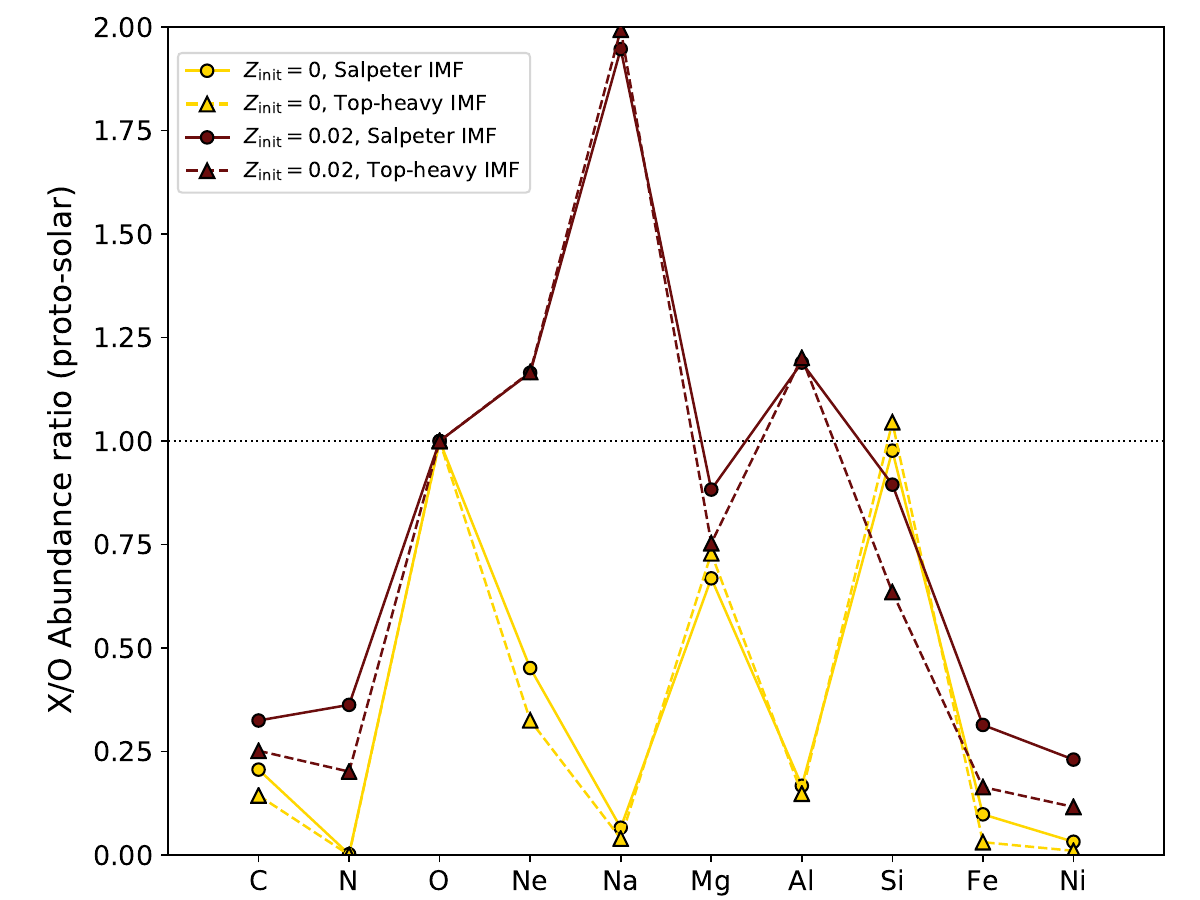}

\caption{X/O abundance ratio yielded by a population of SNcc\cite{Nomoto13} as a function of the initial metallicity of their progenitors (assuming a Salpeter IMF for the latter; left panel) and of the slope of the stellar IMF (assuming the initial metallicity to be either Solar or zero; right panel). As seen in the left panel, the most sensitive indicators of massive stars' initial metallicities are Na/O and Al/O. As seen in the right panel, the most sensitive indicators of the slope of the IMF are Ne/O and/or Mg/O (depending on the initial metallicity). Accurate measurements of these abundance ratios in the ICM allow to constrain these properties in a unique and unbiased way.}

\label{fig:SNcc}
\end{figure*}

\subsection{What are the sources of metals in the Universe?}\label{subsec:yields}

As stated earlier, Fe-peak elements are almost entirely produced during SNIa explosions while $\alpha$-elements predominantly originate from SNcc and their massive star progenitors. Intermediate elements such as Si and S are produced by the above two channels in comparable proportions, whereas C and N are mostly synthesized by intermediate-mass stars during their asymptotic giant branch (AGB) phase\cite{Nomoto13}. Quite remarkably, the physics, the environmental conditions, as well as the stellar initial mass function (IMF) of these stellar populations have a large influence on the relative proportion of chemical elements released by these metal factories in the Universe. Inversely, an accurate estimate of the chemical composition of the ICM, which has emerged through the average imprint of billions of supernovae and AGB stars over Mpc scales, represents a truly unique way of accessing (i) the contribution of these three enrichment channels and (ii) their astrophysical and/or environmental conditions. Abundance ratios in the ICM constitute thus a unique signature that connects the production of metals by stars and supernovae to their enrichment at very large scales. For such investigations, \xrism\ will be limited to the very central parts of relaxed clusters (covering less than 5\% of the total cluster volume), with limited information on possible spatial differences between these ratios. On the contrary, the non-dispersive, high resolution spectroscopy, and wide FoV of \lem\ makes it the perfect mission capable of probing abundance ratios in cluster cores \textit{and} outskirts with unprecedented accuracy---unveiling for the first time the sources of metals in the entire cluster volume. This distinction between cluster core and outskirts is essential because the stellar population of the giant central brightest cluster galaxy (BCG), found in the core of relaxed systems, presumably contributes to the enrichemnt of the central parts of the ICM at a non-negligible degree. Separating these regions, \lem\ will answer three fundamental questions:

\subsubsection{What are the relative contributions of SNcc, SNIa, and AGB stars to metal production?}

The relative contribution of SNcc vs. SNIa in cluster cores has been investigated in previous work\cite{Werner06,dePlaa07,Mernier16b,Simionescu19,Sarkar22} by fitting the abundance pattern (specifically the X/Fe abundance ratios) measured in central cluster regions with a linear combination of SNcc and SNIa nucleosynthesis yield models. Imposing $\mathrm{Fe}_{\mathrm{SNcc}}/\mathrm{Fe} + \mathrm{Fe}_{\mathrm{SNIa}}/\mathrm{Fe} = 1$ , the actual fraction of SNIa (or SNcc) having contributed to the ICM enrichment constitutes the only free parameter of the fit. The systematic uncertainties of many abundance ratios (in particular for $\alpha$-elements) due to the moderate spectral resolution of CCD spectrometers, however, have not allowed us yet to place strong constraints on such relative contribution\cite{deGrandi09}. The case of AGB stars is even more challenging since the N/Fe ratio could be obtained only with the \xmm\ gratings in the peaked, very central parts of a handful of groups and isolated ellipticals\cite{Mao19,Mernier22}. As demonstrated further in Sect.~\ref{sec:LEM}, \lem\ meets all the requirements to measure accurately the C/Fe, N/Fe, O/Fe, Ne/Fe, Na/Fe, Mg/Fe, Al/Fe, Si/Fe, and Ni/Fe ratios all at once in nearby clusters, hence to determine the fraction of AGB, SNcc and SNIa having enriched the Universe.

\subsubsection{Were progenitors of SNcc already rich in metals?} 

The pre-enrichment scenario discussed earlier predicts that stars having enriched the ICM must have had low metallicities before their explosion as supernovae. Quite remarkably, initial metallicities of SNcc progenitors ($Z_\mathrm{init}$) can be traced accurately by their post-explosion abundance pattern. As can be seen in Fig.~\ref{fig:SNcc} (left), at fixed stellar mass nucleosynthesis calculations of SNcc\cite{Nomoto13} exhibiting different $Z_\mathrm{init}$ predict very different abundance patterns. This is particularly true for the Na/O and Al/O ratios, both of which are still unconstrained by current CCD instruments. Whereas \xrism\ should be able to detect the main emission lines of these odd-Z elements and provide rough estimates of their abundances in the brightest core of the most nearby systems, the better energy resolution and larger grasp of \lem\ will clearly be an advantage for this science case. The constraints of \lem\ on the initial metallicity of massive stars having enriched the Universe will also constitute a yet missing piece to solve the puzzle of the ``Fe conundrum'' in clusters. Specifically, the total Fe mass contained in the hot gas of massive haloes is measured with a factor $\sim$2--6 excess compared to what current SNIa can reasonably produce in today's galaxies\cite{Renzini14,Ghizzardi21}. A pre-enrichment scenario, in which metals in the ICM and the stellar population seen in clusters do not have a one-to-one correspondence, seems possible to explain this tension. Such a scenario naturally implies that a low $Z_\mathrm{init}$ for the progenitors of the early SNcc population having enriched the ICM. In addition, stars forming from such metal-poor clouds must have been considerably more massive than now, potentially releasing a very large amount of metals per exploding star. In other words, the initial mass function (IMF) of these early-enriching stars must have been different of (potentially with more massive stars) than seen today in our Galaxy. This aspect is discussed further in the next question.


\subsubsection{What is the shape of the stellar initial mass function and how universal is it?}

This question is still actively debated as the spatial universality of the stellar initial mass function (IMF) may have profound consequences for the evolution of galaxies\cite{Bastian10}. In particular, optical/infrared studies\cite{Smith20} and chemical evolution codes\cite{Yan21} suggest that the slope of the IMF in massive galaxies differs from its canonical shape. As the production of elements by SNcc and AGBs depends on the mass of their progenitors, the average slope of the IMF is also encoded in the chemical composition of the ICM. As seen in Fig.~\ref{fig:SNcc} (right), a Salpeter or top-heavy (i.e. flat) IMF can be clearly distinguished from the Ne/O or Mg/O ratio (depending on the initial metallicity, which is primarily determined by Na/O and Al/O as mentioned above). Comparing the slope of the IMF in cluster cores (where the BCG resides) and in cluster outskirts (enriched by galaxies of lower masses)---as well as with best estimates for our Milky Way from optical/IR surveys---will allow us to conclude this long-standing debate. Besides, \lem\ will solve the Fe conundrum mentioned above---not only by determining the initial metallicity of the enriching stellar population, but also by decisively (dis-) proving the importance of massive stars in the Fe budget of the Universe.

\section{Potential and capabilities of \lem}\label{sec:LEM}

In this Section, we now quantify how and to which extent \lem\ will address the science questions and objectives discussed earlier. Concretely, three different approaches are considered: (i) simple simulated \lem\ spectra based on real estimates from \xmm\ observations of a relaxed nearby cluster (Abell\,3112), (ii) a relaxed cluster simulated in the TNG300-1 run of the IllustrisTNG project and spectrally analyzed through \lem\ instrumental response, and (iii) a stacked \lem\ spectral analysis (based on simple simulated spectra) of protoclusters that will be serendipitously discovered during the \lem\ observational campaign. The motivation for each approach---as well as the questions they aim to address---are detailed in the following subsections. 

\begin{figure}[t]
\centering
\includegraphics[width=0.48\textwidth, trim={0cm 0cm 0cm 0cm},clip]{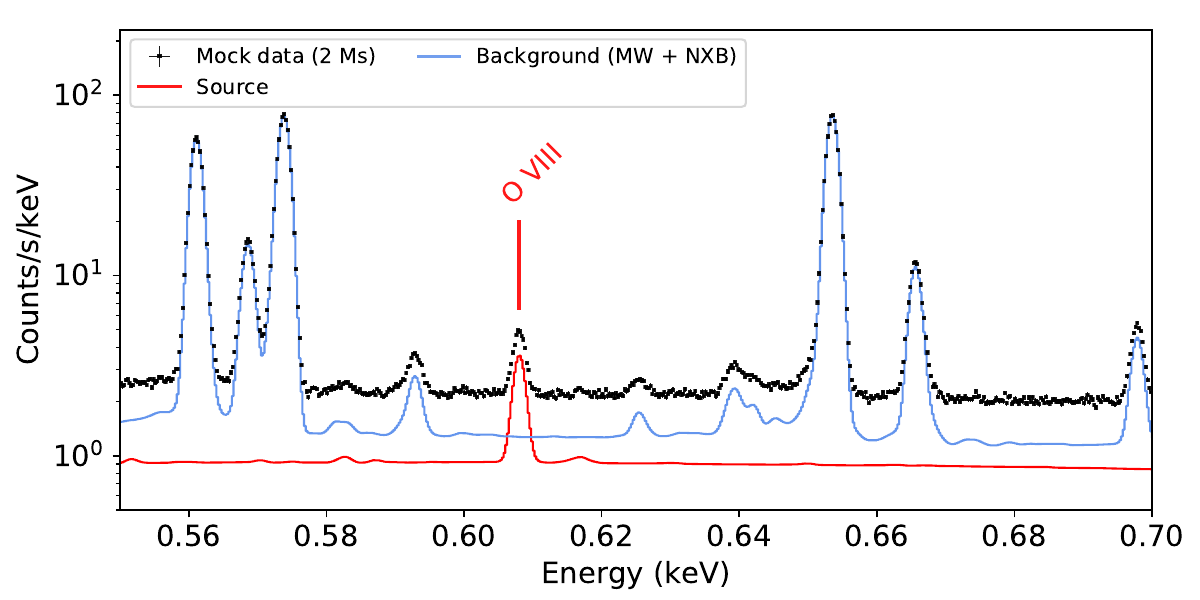} \\
\includegraphics[width=0.48\textwidth, trim={0cm 0cm 0cm 0cm},clip]{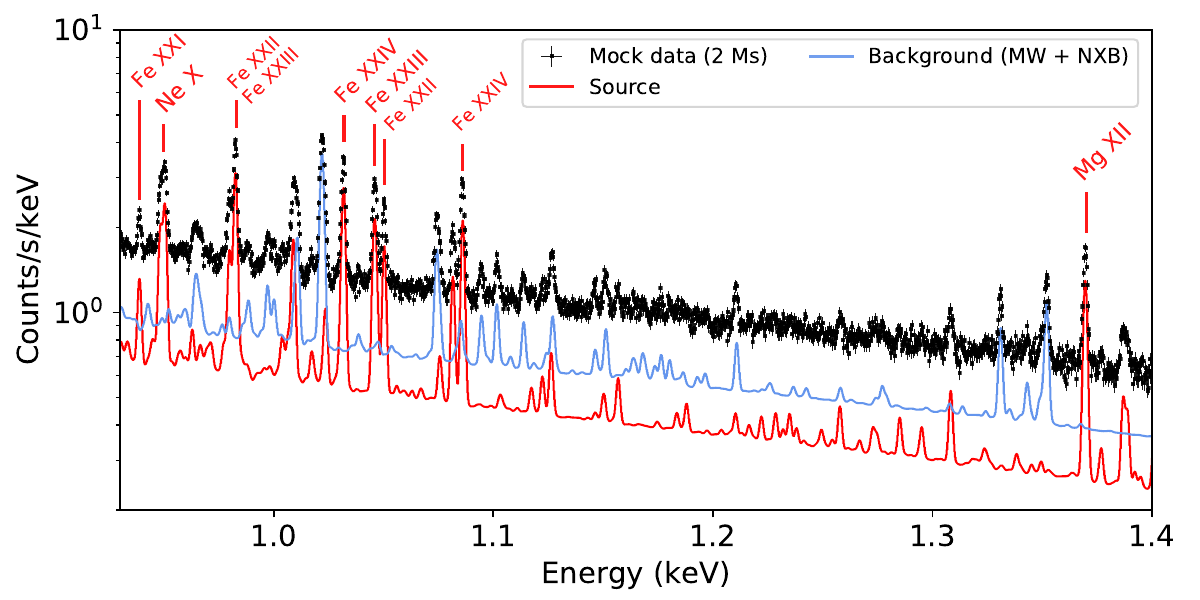}

\caption{Simulated spectrum of a 2~Ms \lem\ observation of the outskirts ($R_{500}$--$R_{200}$) of the galaxy cluster Abell\,3112 ($z \sim 0.075$). Although the dominant source of continuum is that of the background (Milky Way foreground and non-X-ray background), many emission lines of the ICM itself remain clearly detected.}

\label{fig:spectrum}
\end{figure}

\subsection{Abell\,3112}\label{subsec:A3112}

\begin{figure*}[h]
\centering
\includegraphics[width=0.9\textwidth, trim={0cm 0cm 0cm 0cm},clip]{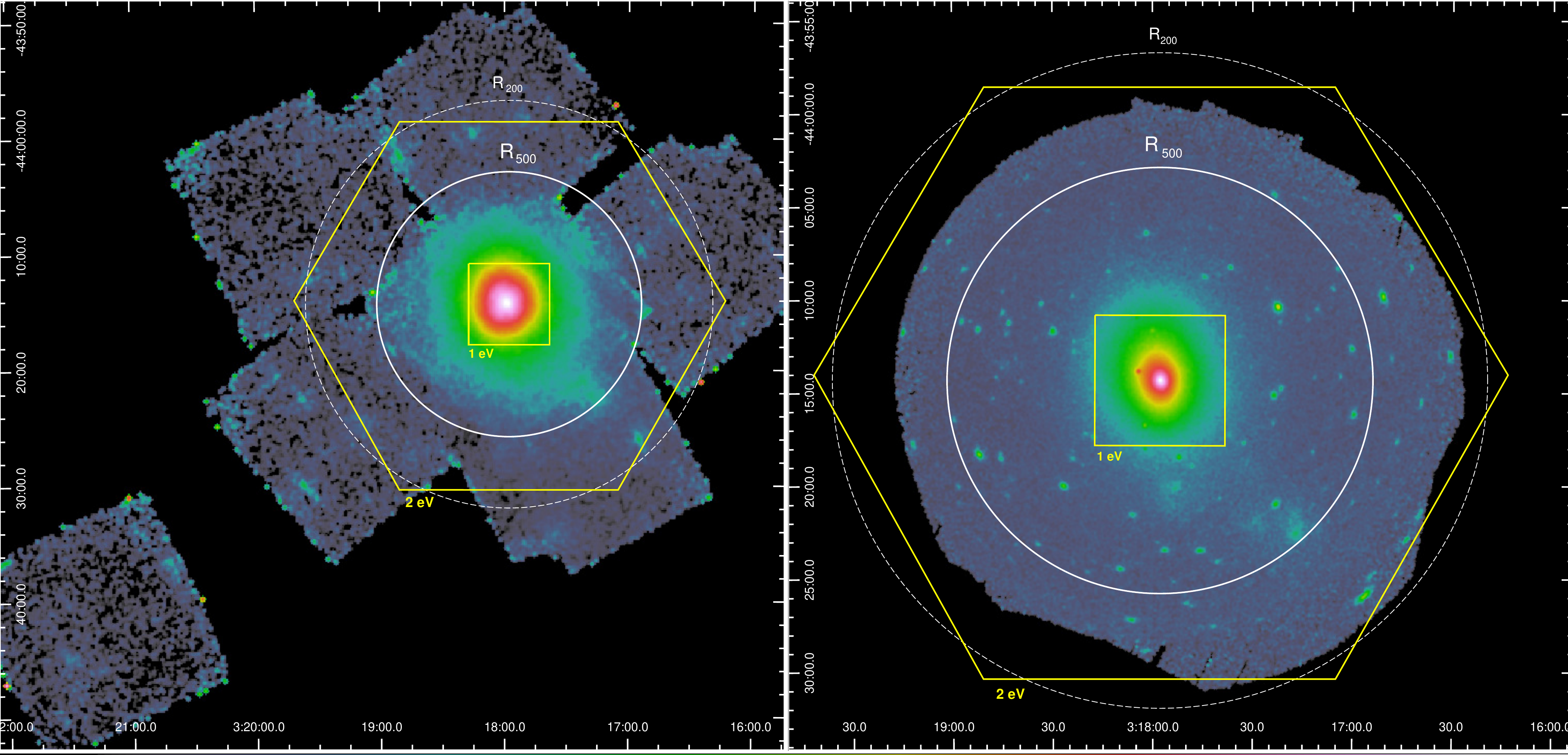}

\caption{\Suzaku/XIS (\textit{left}; 0.5--2~keV) and \xmm/EPIC (\textit{right}; 0.3--2~keV) exposure-corrected mosaic images of the relaxed nearby cluster Abell\,3112. Each image also indicates the $R_{500}$ and $R_{200}$ limits of the cluster. In addition, the \lem\ FoV (yellow square and hexagon in each map) can be easily compared with those from the two CCD instruments.}

\label{fig:image_A3112}
\end{figure*}

Abell\,3112 is a nearby cool-core galaxy cluster that has two remarkable advantages. First, spatial spectroscopy studies have been extensively dedicated to this source using the latest generation of (moderate resolution) X-ray observatories\cite{Bulbul12,Ezer17}, offering thus a coherent picture on the typical measurements and statistics to expect with \lem. Second, this cluster lies in the  $0.07 \lesssim z \lesssim 0.08$ redshift range (precisely at $z = 0.075$), which is optimal to isolate (shifted) spectral lines from the source from those of the (rest-frame) Milky Way halo (Fig.~\ref{fig:spectrum}). Clusters at this redshift also allow to be fully covered by the \lem\ FoV out to at least $R_{500}$, hence to take full advantage of the instrument in one pointing only (Fig.~\ref{fig:image_A3112}).

This first approach is thus based on an azimuthally symmetrical toy model, where the projected emission measure, temperature, and abundance radial profiles are estimated from best-fitting spectral parameters in nine concentric annuli previously investigated with \xmm\cite{Mernier17}. The (single-temperature) best-fitting models obtained from that previous analysis are re-simulated assuming the standard \lem\ response files and 2~Ms of combined observation. Whereas Fe is assumed to trace its best-fitting \xmm\ value, for simplicity we model all other elements to follow the former profile (i.e. X/Fe = 1 everywhere). The mock \lem\ spectra are refitted in XSPEC assuming an absorbed single-temperature plasma (\texttt{vvapec} model). The consequences of this simple modelling on our feasibility are discussed further below. The X-ray background (including the local hot bubble, the Milky Way halo, and the cosmic X-ray background) is assumed to follow the state-of-the-art estimates\cite{McCammon2002,Hickox2006} whereas the detector background is modelled as constant at 1 count/s/keV per LEM FoV, following the instrumental expectations. Following the \lem\ instrumental setup, we also assume an energy resolution of 1~eV and 2~eV, respectively inside and outside 4$'$ of radius. 

Figure~\ref{fig:spectrum} shows the \lem\ spectrum expected for the outermost annulus ($R_{500}$ to $R_{200}$). Quite expectedly, the X-ray background continuum dominates over that of the source. Such a situation would be highly problematic at CCD resolution since the two components would them remain indistinguishable. The exquisite spectral resolution of \lem\ is a game changer for this case: many lines (including O~VIII, Ne~X, Mg~XII, Fe~XXII, and Fe~XXIII) dramatically peak above the background continuum and can be robustly detected\cite{Zhang24}. This ensures, at the very least, accurate measurements of the ICM temperature as well as of relative abundance ratios (e.g. O/Fe, Mg/Fe) as they do not depend directly on the continuum. Even better: the \lem\ all-sky survey\cite{LEMASS} will characterize accurately the spectral behavior of the Milky Way foreground, which will be largely kept under control in our spectral fits. Even though the normalization of the detector background is left unconstrained in our fits, the constraints on the relative abundance ratios \textit{and} absolute abundances remain excellent.

\begin{figure*}
\centering
\includegraphics[width=0.48\textwidth, trim={0cm 0cm 0cm 0cm},clip]{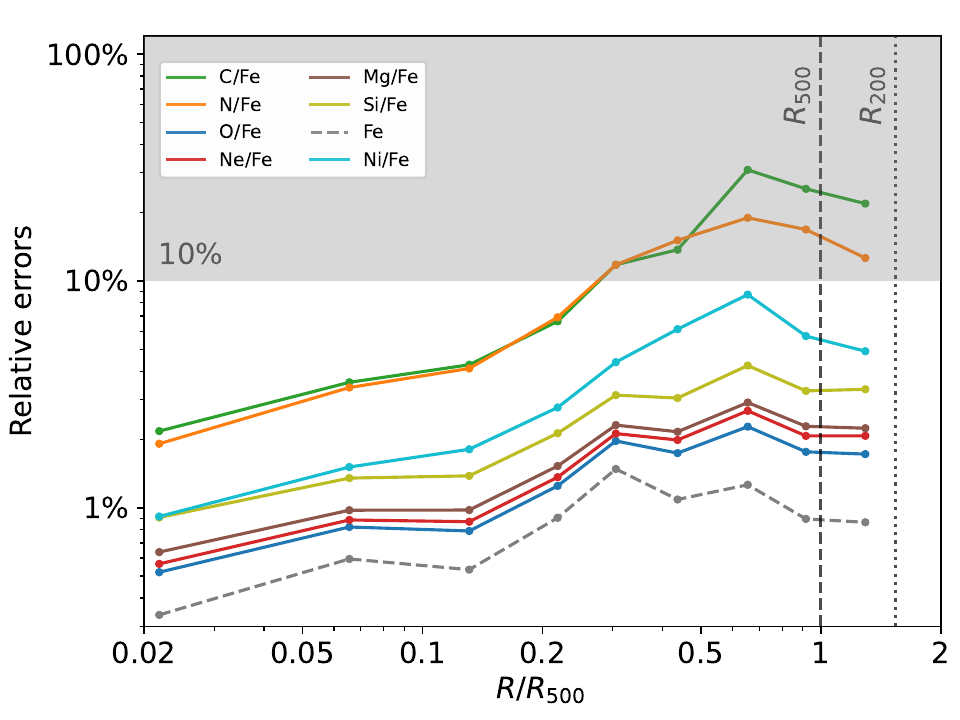}
\includegraphics[width=0.48\textwidth, trim={0cm 0cm 0cm 0cm},clip]{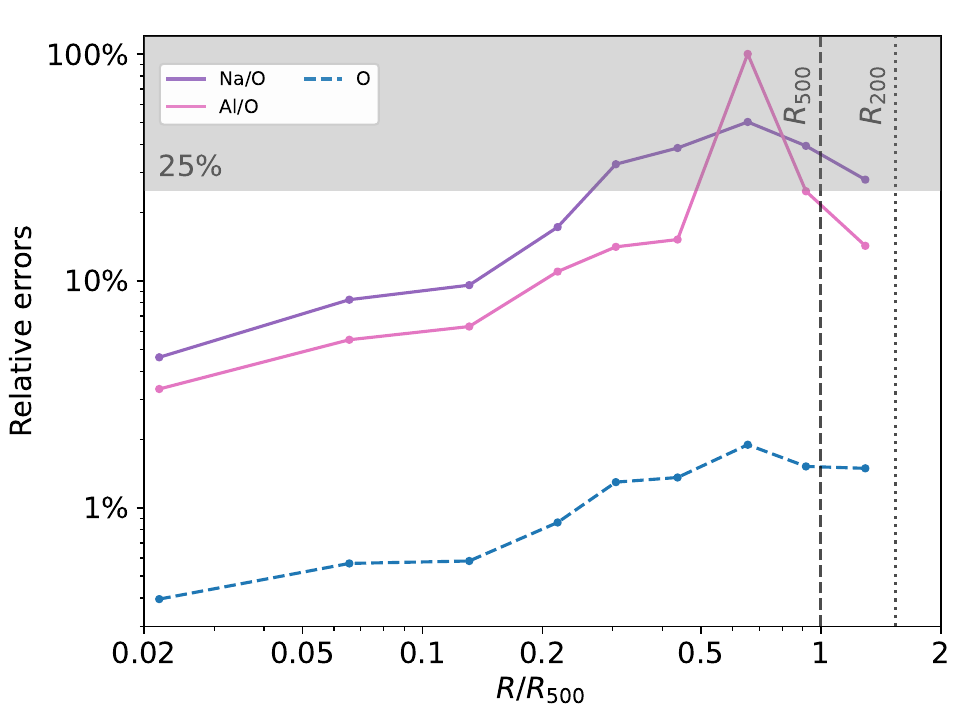} \\
\includegraphics[width=0.48\textwidth, trim={0cm 0cm 0cm 0cm},clip]{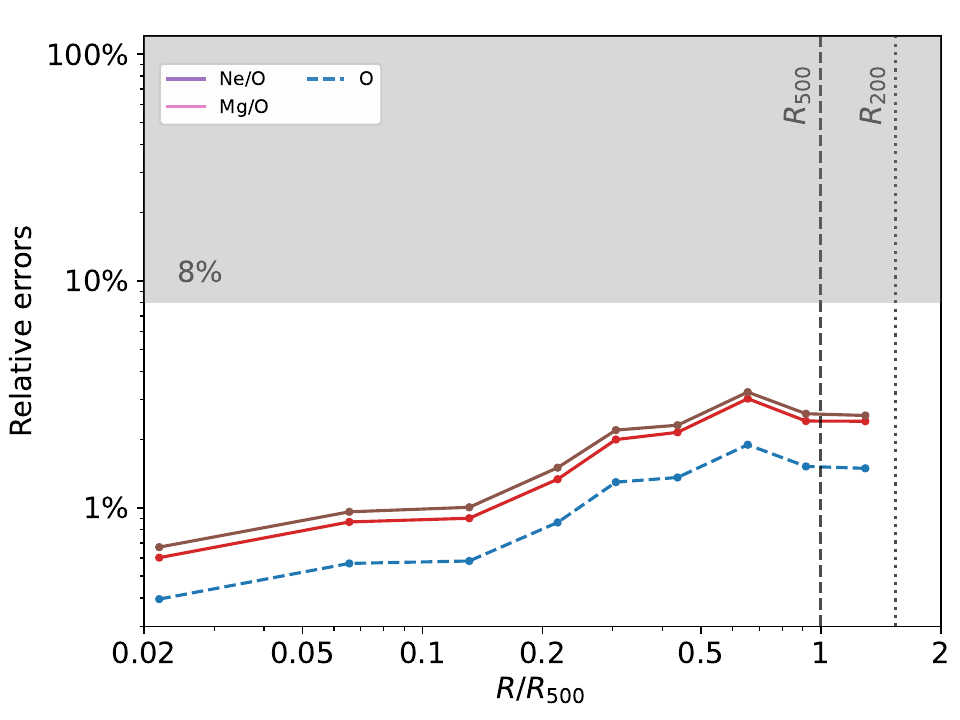}
\includegraphics[width=0.48\textwidth, trim={0cm 0cm 0cm 0cm},clip]{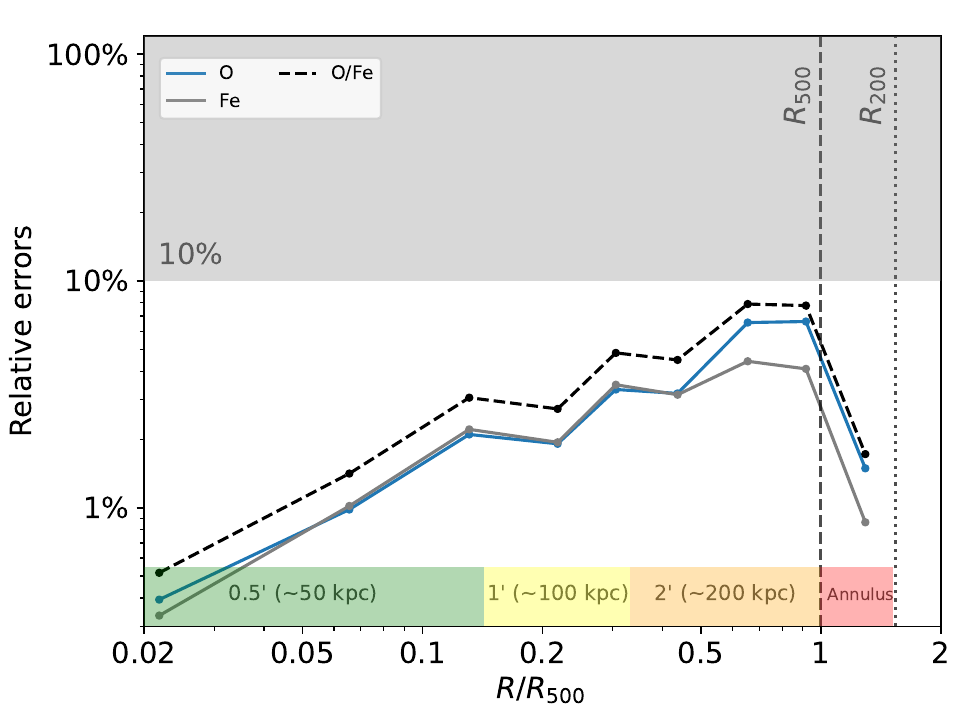}

\caption{Expected precision of key abundance ratios for a cluster like Abell\,3112 ($z \sim 0.075$) as a function of the distance from its core. The white area delimits the region within the precision required to achieve our science goals (written in grey above the limit; see also text).  \textit{Top left:} Fe abundance and X/Fe ratios (derived over concentric annuli), necessary for knowing the radial variation of the relative contribution of AGB stars, SNcc, and SNIa in the ICM. \textit{Top right:} O abundance and X/O ratios (derived over concentric annuli), necessary for knowing the radial variation of the initial metallicity of massive stars that enriched the ICM. \textit{Bottom left:} O abundance and X/O ratios (derived over concentric annuli), necessary for knowing the radial variation of the slope of the IMF of the enriching stellar population. \textit{Bottom right:} O and Fe abundance and the O/Fe ratio (derived over individual spatial cells of various sizes), necessary for deriving the spatial variation of relative contribution of SNcc and SNIa in the ICM. A simple, close to optimized configuration with cell sizes increasing with radius is proposed (color bar at the bottom of the panel).}

\label{fig:radial_A3112}
\end{figure*}

The relative errors expected for these abundances / ratios at all radii are illustrated further in Fig.~\ref{fig:radial_A3112}. The top left panel shows errors for the Fe abundance as well as for seven ratios that are essential to determine the relative contribution of AGB stars, SNcc, and SNIa to the enrichment as a function of the radial distance. Following simple estimates using currently favored nucleosynthesis models, uncertainties of 10\% on each of these ratios (shown in grey on the figure), give approximately 10\% accuracy on the relative fraction of SNcc and of AGB stars contributing to the enrichment. As seen on the same figure, the enriching fraction of SNcc will be, thus comfortably, constrained within $\lesssim$10\%, even beyond $R_{500}$. On the other hand, AGB stars will have their relative contribution known with 10\% accuracy at $0.3\,R_{500}$, which is clearly outside of the central abundance peak, thus outside of the influence zone of the BCG\cite{Mernier22b}. The top right panel shows the expected uncertainties for the O abundance as well as for the Na/O and Al/O ratios. These two ratios are the most sensitive tracers of the initial metallicity of SNcc progenitors. Ignoring the other elements, to remain conservative, we estimate that an uncertainty of $\sim$25\% on each of these two ratios is enough to discriminate (with $3\,\sigma$ significance) two adjacent SNcc yield models in the parameter space $Z_{init} \in [0, 0.001, 0.004, 0.008, 0.02]$\cite{Nomoto13}. Initial metallicities of massive stars can be constrained accurately (and compared radially) out to $0.3~R_{500}$. Beyond that radius, acceptable constraints will be reasonably achieved as long as the analyzed spectra are extracted from broader annuli. The bottom left panel focuses on the Ne/O and Mg/O ratios, which are key tracers of the slope of the IMF. Uncertainties of 8\% on each of these two ratios are enough to discriminate between a Salpeter IMF and a top-heavy IMF at $3\,\sigma$ confidence. Remarkably, the tight constraints obtained from the O, Ne, and Mg abundances (whose main emission lines have large equivalent widths) makes this goal easily achievable out to $R_{200}$. We note that similar conclusions have been reached with the \athena/X-IFU instrument in a sample of four clusters---at least when analyzed over their entire volume out to $R_{500}$\cite{Mernier20}. Last but not least, the bottom right panel quantifies the spatial accuracy over which 2D maps of O, Fe, and of the O/Fe ratio will be obtained. Such maps will be invaluable to address the abundance patchiness of the ICM and to determine the relative contribution of various astrophysical mechanisms of metal enrichment outside galaxies (i.e. early AGN feedback, stellar feedback, ram-pressure stripping, galaxy-galaxy interactions, etc.). The inner $0.15\,R_{500}$ is solely limited by the PSF of the instrument and can easily reach scales below 50~kpc (assuming $z = 0.075$). Map cells covering individual sizes of 100~kpc and 200~kpc will be obtained within, respectively, $0.3\,R_{500}$ and $R_{500}$. Beyond that radius, 2D maps will become challenging and spatial analysis will be performed over an entire outermost annulus (possibly divided in a couple of sectors).

\begin{figure}[t]
\centering
\includegraphics[width=0.48\textwidth, trim={0.5cm 0cm 0cm 0.7cm},clip]{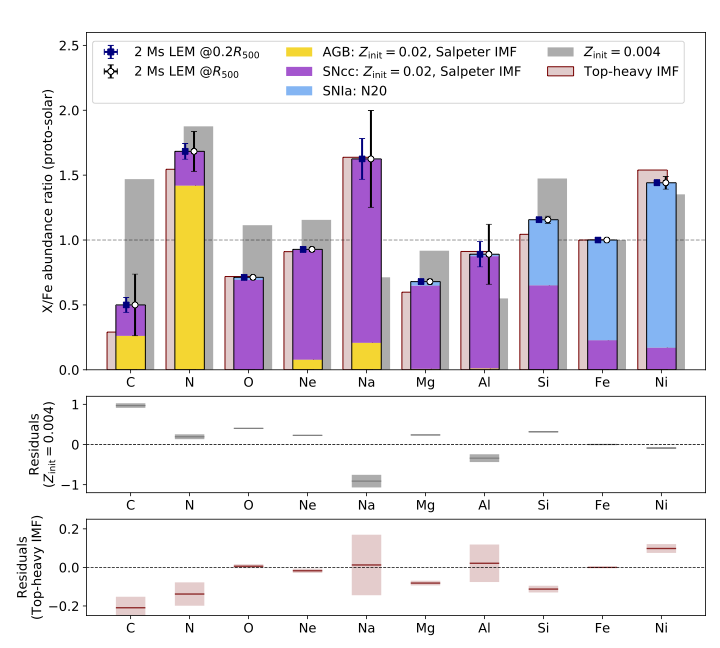}

\caption{Simulated chemical composition of the ICM (X/Fe ratios; i.e. Fe is set to 1 Solar) originating from an enrichment with contributions from a mix of AGBs\cite{Karakas10}, SNcc\cite{Nomoto13}, and SNIa\cite{Seitenzahl13}. The yields adopted here are among the most realistic so far\cite{Simionescu19}. Measured with 2 Ms of \lem\ exposure in a typical z = 0.075 cluster at $0.2\,R_{500}$ (filled error bars) and at $R_{500}$ (empty error bars), this baseline model (histograms) also shows the relative contribution of AGBs, SNcc, and SNIa for each element. Solid lines show alternative predictions of the abundance pattern for a change of either the initial metallicity of AGB and SNcc progenitors (greay shadow) or the slope of the IMF (red shadow). The significant residuals of the measurements compared with the latter two alternative models (two lower panels, shown for the inner $0.2\,R_{500}$) emphasize the robustness at which various progenitor initial metallicities and IMF shapes will be either favored or ruled out.}

\label{fig:ratios}
\end{figure}

A more revealing view of the ground-breaking science offered by \lem\ on constraining the yields of the enriching stellar populations is presented in Fig.~\ref{fig:ratios}. Assuming a combination of three state-of-the-art yield models (with $Z_\mathrm{init} = 0.02$ and a Salpeter IMF for the AGB stars\cite{Karakas10} and the SNcc\cite{Nomoto13} models---both integrated over a Salpeter IMF, as well as a delayed-detonation explosion ignited by 20 spots for the SNIa model\cite{Seitenzahl13}) which provides a near-Solar abundance pattern\cite{Mernier18b,Simionescu19}, the histograms show the relative contribution from AGB stars, SNcc, and SNIa to each element. Assuming the best-fit \lem\ abundance measurements to reproduce our input yields with no bias\footnote{Based on previous dedicated feasibility studies for the \athena/X-IFU instrument\cite{Cucchetti18,Mernier20}, this assumption seems valid for high-resolution X-ray spectroscopy in general.}, we show the statistical uncertainties associated with our measurements obtained in annuli centered around $0.2\,R_{500}$ and $R_{500}$. For comparison, the grey and red shadows illustrate the change in the abundance pattern when assuming, respectively, either another initial metallicity for SNcc progenitors or a top-heavy IMF. The two lower panels show the residuals between our measurements and such alternative models. The remarkably small uncertainties on the abundance ratios that will be measured by \lem\ demonstrate further that the two goals mentioned above (i.e. constraining the initial metallicity of enriching massive stars and the slope of the stellar IMF) will be comfortably achieved by the mission.

While \lem\ will thus constrain the most realistic yield models available so far and deepen our understanding of the ICM \textit{and} stellar physics, one should keep in mind that current nucleosynthesis models have their own uncertainties (related to e.g. numerical effects, initial conditions and assumptions, 2D vs. 3D explosion modelling). Nevertheless, considerable progress is expected in the field over the next decade,  which will be highly beneficial to \lem. In turn, it is very likely that the microcalorimeter era we are entering with \xrism\ (and later \lem) will boost theoretical research on stellar nucleosynthesis and lead to a large number of physically accurate yield models.

The advantages of the above-described exercise are its relative simplicity to provide proper cluster estimates as realistically observed, as well as its flexibility on the observational setup (e.g. tracing rare elements like Na and Al, which are not embedded yet in all cosmological simulations). Its main drawback, however, is that the assumptions of our toy model are sometimes over-simplistic, particularly regarding the temperature structure of the gas. Assumed to be single-temperature here, is it well known that a multi-phase gas can bias abundance estimates, at least at CCD resolution spectroscopy\cite{Gastaldello21}. Encouragingly, we have reproduced the above approach assuming two temperatures as well (i.e. \texttt{vvapec+vvapec}), with no significant bias on the abundance measurements. Such biases, however, might emerge when more complicated temperature structures are at play, for instance those seen in cosmological simulations. The latter also account for (small-scale and large-scale) gas motions, which may potentially affect the abundances as well if models are not fitted properly. An alternative, simulation-based approach is thus proposed in the next section. This said, the main effect of abundance biases are expected to be on abundance measurements themselves, not their associated (statistical) uncertainties. Our above measurements remain thus valid in term of feasibility if one sees them as estimates of \textit{precision} rather than accuracy (i.e. the relative errors as shown in Fig.~\ref{fig:radial_A3112}).

\begin{figure*}[h]
\centering
\includegraphics[width=0.95\textwidth, trim={1cm 0cm 0.4cm 0cm},clip]{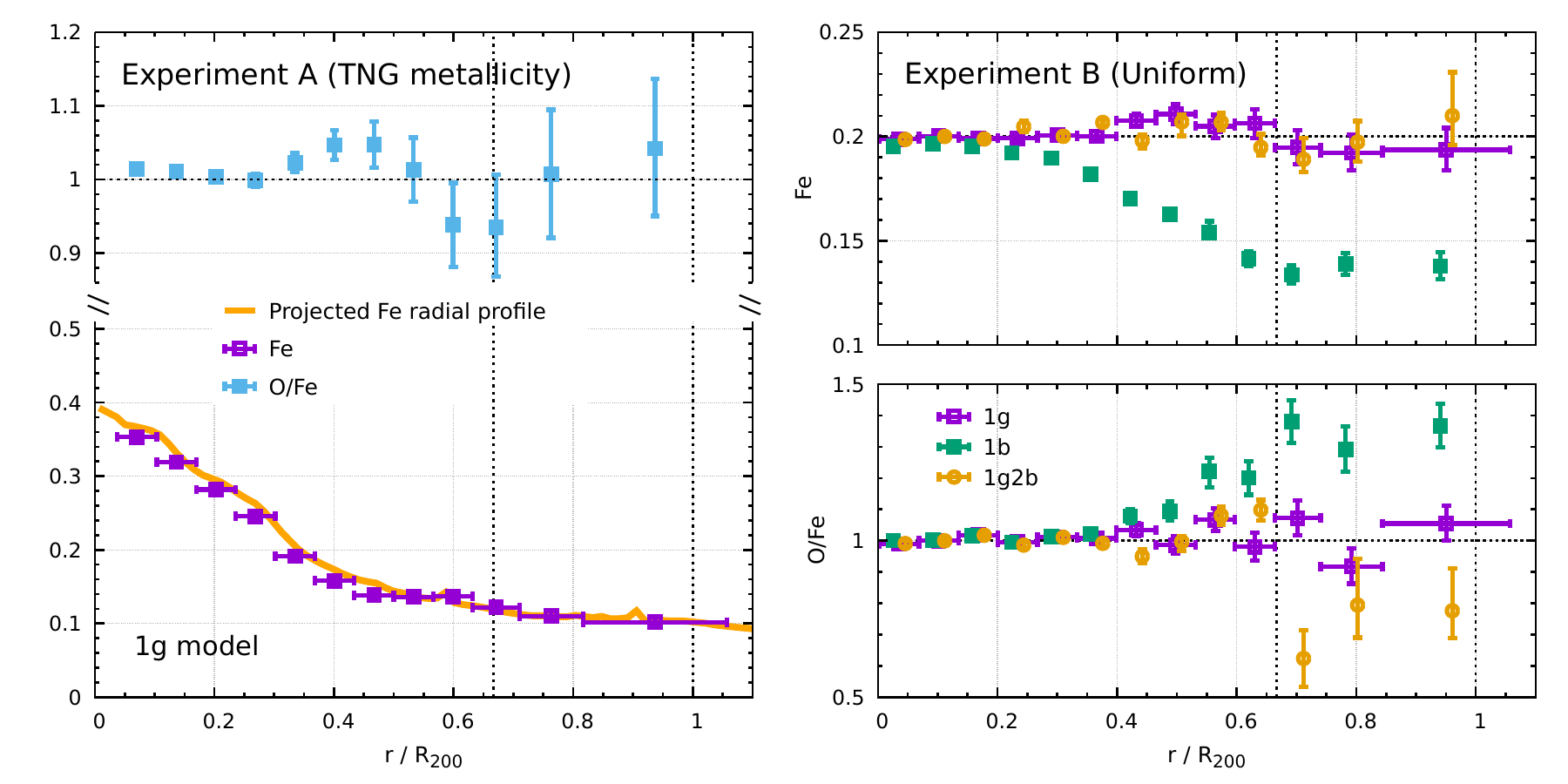}

\caption{\lem\ capability of measuring abundance radial profiles up to $R_{200}$ in one of the most relaxed TNG galaxy clusters at $z = 0.08$. \textit{Left:} an experiment (A) using TNG metallicity distribution and uniform metal abundance ratio (unity). The orange curve shows the X-ray-weighted simulation prediction of the projected metallicity radial profile, where we include only the gas particles with $kT > 1$~keV to exclude the impacts of cold clumps. We can well recover the metal profiles similar to the uniform metallicity case shown on the left (purple points). \textit{Right:} simplified case (B) in which we compare various plasma models of various complexity to recover Fe abundances and O/Fe from the \lem\ 2~eV resolution mock spectra while assuming uniform metal abundances of 0.2~Solar. By excising cold clumps and accounting for the Gaussian distribution of the ICM temperature (i.e. the 1g model), one can reproduce the abundances accurately. Adapted from Zhang et al. (submitted)\cite{Zhang24}.}

\label{fig:TNG}
\end{figure*}

\subsection{Simulated nearby cluster (TNG)}\label{subsec:TNG}

As discussed in the previous section, a feasibility study based on a simulated cluster nicely complements our estimate above based on a real, observed cluster. Though less suitable for tracing rare metals, the main advantage of using mock observations from hydrodynamical simulations is a realistic treatment of complicated effects, e.g. the temperature and velocity structure of the gas, which may impact the abundance measurements.

For this exercise, we use a simulated cluster from the TNG300-1 cosmological simulation as part of the IllustrisTNG project\cite{Nelson18,Pillepich18,Springel18,Naiman18,Marinacci18}. A detailed description of this simulated system and of the capabilities of \lem\ to map its velocity structure are provided in Zhang et al. (submitted)\cite{Zhang24}. This selected, intermediate-mass cluster (ID180645, $M_{200} = 2.8 \times 10^{14}~M_\odot$) is fully relaxed and is set to a redshift of 0.08 for an optimal separation between the source and foreground emission lines. Mock \lem\ observations are then obtained by sampling synthetic photons using \texttt{pyXSIM}\footnote{\href{https://hea-www.cfa.harvard.edu/~jzuhone/pyxsim}{https://hea-www.cfa.harvard.edu/$\sim$jzuhone/pyxsim}} (assuming an \texttt{apec} model), which are then projected and convolved with the \lem\ instrumental response using \texttt{SOXS}\footnote{\href{https://hea-www.cfa.harvard.edu/~jzuhone/soxs}{https://hea-www.cfa.harvard.edu/soxs}} assuming 2~Ms of total exposure. The background is modelled and implemented in a similar fashion as described above (though as fixed parameters in the fit). 

We consider two separate experiments: (A) the distribution of metallicity (in practice, mostly traced by the Fe abundance) is directly extracted from the simulation (as found to be spatially non-uniform); and (B) the metallicity is locked to 0.2~Solar (in units of Anders \& Grevesse 1989\cite{Anders89}, which corresponds to the 0.3~Solar floor in units of Lodders et al. 2009\cite{Lodders09} discussed above) everywhere in the cluster. In both cases, all the X/Fe ratios are assumed to be 1. While experiment A acknowledges that TNG-simulated clusters are predicted to be less enriched than what current observations suggest\cite{Vogelsberger18} (despite a correction factor to extrapolate the TNG300-1 results from the higher resolution TNG100-1 results\cite{Zhang24}), it has the advantage of being seen as a ``worst-case scenario'' in which line emissivities in cluster outskirts are lower than expected. Experiment B, on the other hand, reflects abundances as currently observed in the outskirts of massive clusters with CCD resolution. Spectra from these two experiments are extracted over concentric annuli and then fitted with appropriate plasma models in XSPEC. Three models are considered, all accounting for line broadening due to bulk gas motions: 
\begin{enumerate}
    \item a (single-temperature) \texttt{bvapec} model (``1b'');
    \item a multi-temperature model consisting of a variant \texttt{bvapec} whose temperature is distributed as a Gaussian function (``1g'');
    \item a multi-temperature model composed of the same Gaussian-like \texttt{bvapec} component as in 2., as well as two (single-temperature) \texttt{bapec} components (``1g2b'').
\end{enumerate}
The ICM is known to become clumpy and multi-phase, particularly in the outskirts. Our mock data from the TNG-simulated cluster show that \lem\ will allow to resolve cool gas clumps and exclude them from spectral analysis\cite{Zhang24}. The former two models assume that this procedure has been applied. The third model instead aims to account for the clumps spectrally (through the addition of the 2b components). 


Figure~\ref{fig:TNG} shows the measured \lem\ radial profiles of these two experiments. Whereas a single-temperature approach clearly biases the Fe and O/Fe estimates (see the right panel for experiment B), we note that the 1g model successfully recovers the input abundance profiles after clump regions have been excluded. The statistical uncertainties depend on the number (and thickness) of extracted annuli; nevertheless this exercise demonstrates that \lem\ will be able to provide measurements with less than 10~\% bias through the entire cluster volume---with the notable caveat that a limited number of counts inevitably implies a non-negligible risk of obtaining a few (randomly distributed) scattered measurements (on the order of $\lesssim$20--30\% in the outermost annuli). This will allow to answer the key science questions developed in Sect.~\ref{sec:questions}.

\begin{figure*}[t]
\centering
\includegraphics[width=0.46\textwidth, trim={0.5cm 0.5cm 1cm 0.5cm}]{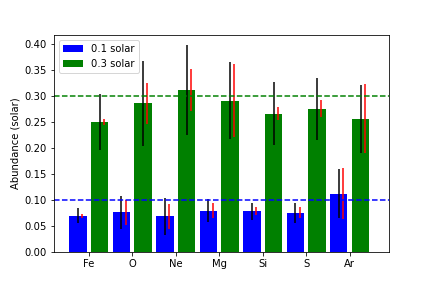}
\includegraphics[width=0.46\textwidth, trim={0.5cm 0.5cm 1cm 0.5cm}]{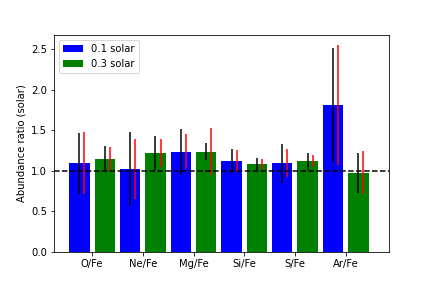}

\caption{Results of the stacking analysis of 15 typical protoclusters at $z\sim2$ that are serendipitously detected by \lem. \textit{Left:} The median of the best-fit abundances for Fe, O, Ne, Mg, Si, S, Ar of 50 realizations are shown in the solid bars. Black bars indicate the standard deviation in the best-fit abundances of the 50 realizations. Typical $1\sigma$ statistical uncertainties of a realization are marked in red. Their true abundances are shown in horizontal dashed lines. Green and blue data correspond to the true abundances of 0.3 and 0.1~Solar, respectively. We will be able to measure the level of metal abundance of nascent ICM in protoclusters and discern whether it is consistent with that of nearby clusters. \textit{Right:} Similar to the left panel but for abundance ratios of O/Fe, Ne/Fe, Mg/Fe, Si/Fe, S/Fe, Ar/Fe. The input (true) abundance ratios are the Solar abundance ratio as shown in the horizontal black dashed line. The green and blue data correspond to input metal abundances of 0.3 and 0.1~Solar, respectively, for the {\tt apec} model. We will be able to measure the $\alpha/$Fe abundance ratios of the ICM in protoclusters in some cases.}

\label{fig:stack}
\end{figure*}

\subsection{Serendipitous detection of protoclusters}\label{subsec:protoclusters}

The \lem\ all sky survey\cite{LEMASS} and deep pointed observations that are driven by other scientific objectives will allow sources, particularly at high redshift, to be serendipitously detected. We will utilize these observations to probe the metal content of the nascent ICM in protoclusters at $z \sim 2$. Cosmological simulations give a prediction of the number density of massive structures at any given redshift. We expect 1--2 protoclusters at $z\sim2$ with a mass of 3--10$\times10^{13}~M_\odot$ falling on each \lem\ field. 
Among various observing campaigns, we expect 20 \lem\ pointed observations planned with $\sim$0.5~Ms each.
From a $\sim$0.5~Ms \lem\ observation with a 2\,eV spectrum resolution, we assume a conservative detection rate of 50\% for such protoclusters. 
This will allow $\sim15$ protoclusters at $z\sim2$ to be detected by \lem\ with a total exposure time of 10~Ms, which essentially ``comes for free". Future optical or sub-mm surveys or followup observations may further improve the detection rate of the protoclusters of \lem\ fields.


The assumed thermal properties of these 15 protoclusters are largely based on those of the spiderweb protocluster\cite{Tozzi22}. Their temperatures are randomly chosen from a uniform distribution between 1--3\,keV. Their ICM flux in 0.5--2.0\,keV is randomly chosen from a Gaussian distribution with a mean of $1.7\times10^{-15}$~erg\,s$^{-1}$\,cm$^{-2}$ and a standard deviation of $3\times10^{-16}$~erg\,s$^{-1}$\,cm$^{-2}$. Protocluster observations may unavoidably contain emission from AGN, for which we assume a power-law model. Its photon index is randomly generated from a Gaussian distribution with a mean of 1.8 and a standard deviation of 0.2. Its luminosity in the 2--10\,keV band is randomly chosen from a uniform distribution between 10$^{43-45}$~erg\,s$^{-1}$.
In addition to the emission from protoclusters, we take into account the instrumental background as well as various sources of astrophysical backgrounds as described earlier (Milky Way, local hot bubble, and cosmic X-ray background). Their parameter values are fixed during our analysis as the large FoV of \lem\ would allow a robust background estimate for high redshift sources. \lem\ response files for a spectral resolution of 2~eV were used to generate the synthetic spectra. 
We consider two cases for the input metal abundance at 0.1 and 0.3~Solar,
as representative of the two ``extreme'' scenarios (significant chemical evolution and no chemical evolution, respectively). The input X/Fe abundance ratios were all set at the Solar abundance ratio. 
For each of the two cases, we generated 100 synthetic spectra from the aforementioned parameter distributions. We randomly selected 15 spectra out of the 100 spectra and we jointly fit these 15 spectra with a thermal model plus a power-law model absorbed by the Galactic foreground in XSPEC: {\tt phabs$\times$(vapec+pow)}. We repeated this process 50 times for each case. We obtained the median and standard deviation of the best-fit values of the 50 realizations, as well as the typical statistical uncertainties of a realization.  
 As shown in Fig.~\ref{fig:stack}, we can distinguish between the two above scenarios of various degrees of enrichment in the nascent ICM in protoclusters. This will allow us to determine whether there is an evolution from this era to that of mature clusters ($\sim$0.3~Solar). It also allows us to determine some of the abundance ratios, in order to put constraints on the relative contribution of SNIa and SNcc yields when the Universe was only 3 Gyr old. 


Massive galaxies with enriched interstellar / circumgalactic medium are being discovered at high redshifts, particularly with JWST\cite{Peng23, Zhang23}. However, such studies typically trace $\alpha$ elements such as O, synthesized in SNcc, often with less quantitative constraints on the abundances due to the more complicated ionizing state of the gas. Our work, on the other hand, can provide unique insights into the SNIa enrichment at such high redshifts, and will allow us to probe with excellent quantitative constraints the origin of metals in the early universe.

As a word of caution, we note that the present exercise is based on simplistic assumptions, as well as on a limited number of realizations. The above predictions (in particular the best-fitting values of Fig.~\ref{fig:stack}) should thus be seen as first-order estimates only. A complete feasibility study would require more realizations of our assumed parameters and proper consideration of the potential impact from the main sources targeted in each \lem\ pointing.

\section{Summary and observing strategies}\label{sec:conclusion}

This White Paper is the outcome of many months of discussions and efforts among a significant fraction of the X-ray cluster community to (i) (re-) formulate today's most urgent questions on the enrichment of the Universe in its hot gas phase and (ii) evaluate whether (and to which extent) \lem\ will be able to answer them. These questions can be summarized as:

\begin{enumerate}
    \item \textbf{When did galaxies enrich the ICM?} The chemical history of the ICM left several key signatures that are directly accessible to observations, in protoclusters at high redshift but also in the outskirts of nearby systems. Whereas most of our efforts of these last decades focused on Fe, accessing and mapping the abundance of other elements constitutes a decisive step in our quest of obtaining a full picture of large-scale enrichment and its evolution. Such a global picture can be reached if---and only if---these different angles of attack (outskirts of nearby clusters, protoclusters, Fe abundance, $\alpha$-elements abundances, etc.) are deeply investigated in synergy.
    
    \item \textbf{What are the sources of metals in the Universe?} Together with dark matter and dark energy, this is one of the rare questions in modern physics that connect directly sub-atomic scales (nucleosynthesis) and cosmic scales (spread of metals in galaxy clusters). The simplicity of the emission processes at play in the ICM offers us the unique opportunity to measure abundances more accurately than in any other astrophysical system (even our own Solar System). This way, we can understand the physical and environmental conditions of the bulk of stars and supernovae of the Universe by investigating their direct imprint in clusters of galaxies. This requires an instrument capable of covering the entire volume of bright, nearby clusters at high spectral resolution. 
\end{enumerate}

Quite remarkably, the \lem\ capabilities offer us a unique opportunity to answer these two questions at once. In this work, we have shown that a deep (2~Ms) observing campaign of a relaxed cluster at $z \simeq 0.07$--0.08, in synergy with serendipitously discovered protoclusters in wide-field surveys, will allow unprecedented mapping of no less than 10 chemical elements. These key observables will provide unique constraints on the chemical history of clusters and on the stellar populations responsible for it. 

While these two observing strategies will answer most of the questions discussed in Sect.~\ref{sec:questions}, the universality of the Fe enrichment (Sect~\ref{subsec:when}) can be achieved only when comparing the outskirts of clusters with that of lower-mass galaxy groups. Accounting for the typical uncertainties obtained in this work, the loss of overall X-ray brightness in groups is (at least partly) compensated by a higher line emissivity in the soft band due to the cooler temperature of the intragroup medium\cite{Reiprich02}. Therefore, as a second step, 2 additional Ms will be devoted to a sample of four other systems---two groups and two massive clusters. This last investigation will shed light on the variation of abundance distributions between similar systems (to confirm that, at fixed mass galaxy clusters tend to be quite similar to one another), but also with system mass. This dual comparison will be essential to determine (or rule out) the universality of the enrichment of the Universe at large scales. Similar Fe, O, and O/Fe profiles would provide decisive evidence toward the pre-enrichment scenario discussed earlier, while other outcomes would redefine our complete view on enrichment at and beyond galaxy scales.

We also stress that beside the aspects discussed in the present study, \lem\ will be able to answer many other questions relevant to chemical enrichment. For instance, the large FoV of the instrument will ensure an excellent coverage of the core (and intermediate regions) of the most nearby systems such as Virgo, Perseus, Centaurus, or even Coma. During the Guest Observer (GO) phase of the mission, \lem\ will thus be able to map metals in these systems to understand how (efficiently) metals are mixed under the action of central AGN jets and X-ray cavities, sloshing gas motions, cold fronts, and even cluster mergers. Since metals may also be seen as passive gas tracers, such studies are highly relevant to our understanding of the ICM enrichment but also on other key properties of the ICM, such as its kinematics, thermodynamics, assembly history, or even viscosity.

As a final note, \xrism\ will undoubtedly be a pathfinder for high-resolution X-ray spectroscopy (allowing the first major results in the cores of nearby systems and considerable improvement of our current spectral models and atomic codes) while the unprecedented effective area of \textit{NewAthena} will allow us to probe the chemical composition of \textit{individual} protoclusters. These two missions will also allow accurate X-ray spectroscopy beyond 2~keV, necessary to probe critical elements such as Ca, Ar, Cr, and Mn. The science addressed by \lem\ on this topic will thus remarkably complement those addressed by \xrism\ and \textit{NewAthena}, offering a bright future for this field (and astrophysics in general) over the next decade(s).


\small


\vspace{-6mm}
\parindent=0cm
\baselineskip=12pt

\begin{thebibliography}{10}

\bibitem{Biffi18}
Biffi, V., Mernier, F., \& Medvedev, P.\ 2018, \ssr, 214, 123.

\bibitem{Mernier18c}
Mernier, F., Biffi, V., Yamaguchi, H., Medvedev, P., Simionescu, A. et al.\ 2018, \ssr, 214, 129.

\bibitem{Mernier22b}
Mernier, F., \& Biffi, V.\ 2022, Handbook of X-ray and Gamma-ray Astrophysics, 12.

\bibitem{Nelson18}
Nelson, D., Pillepich, A., Springel, V., Weinberger, R., Hernquist, L. et al.\ 2018, \mnras, 475, 624.

\bibitem{Pillepich18}
Pillepich, A., Nelson, D., Hernquist, L., Springel, V., Pakmor, R. et al.\ 2018, \mnras, 475, 648.

\bibitem{Springel18}
Springel, V., Pakmor, R., Pillepich, A., Weinberger, R., Nelson, D. et al.\ 2018, \mnras, 475, 676.

\bibitem{Naiman18}
Naiman, J. P., Pillepich, A., Springel, V., Ramirez-Ruiz, E., Torrey, P. et al., S.\ 2018, \mnras, 477, 1206.

\bibitem{Marinacci18}
Marinacci, F., Vogelsberger, M., Pakmor, R., Torrey, P., Springel, V. et al.\ 2018, \mnras, 480, 5113.

\bibitem{Mernier18b}
Mernier, F., Werner, N., de Plaa, J., Kaastra, J. S., Raassen, A. J. J. et al.\ 2018, \mnras, 480, L95.

\bibitem{Simionescu19}
Simionescu, A., Nakashima, S., Yamaguchi, H., Matsushita, K., Mernier, F. et al.\ 2019, \mnras, 483, 1701.

\bibitem{Werner08}
Werner, N., Durret, F., Ohashi, T., Schindler, S., \& Wiersma, R. P. C.\ 2008, \ssr, 134, 337.

\bibitem{XRISM20}
XRISM Science Team\ 2020, arXiv e-prints, arXiv:2003.04962.

\bibitem{Barret23}
Barret, D., Albouys, V., den Herder, J.-W., Piro, L., Cappi, M., et al.\ 2023, Experimental Astronomy, 55, 373.

\bibitem{Kraft22}
Kraft, R., Markevitch, M., Kilbourne, C., et al.\ 2022, White Paper, arXiv:2211.09827.

\bibitem{Nelson23}
Nelson, D., Byrohl, C., Ogorzalek, A., Markevitch, M., Khabibullin, I. et al.\ 2023, \mnras, 522, 3665.

\bibitem{Truong23}
Truong, N., Pillepich, A., Nelson, D., Bogd\'{a}n, \'{A}., Schellenberger, G. et al.\ 2023, \mnras, 525, 1976.

\bibitem{Schellenberger23}
Schellenberger, G., Bogd\'{a}n, \'{A}., ZuHone, J. A., Oppenheimer, B. D., Truong, N. et al.\ 2023, \apj\ (subm.), arXiv:2307.01259.

\bibitem{ZuHone23}
ZuHone, J. A., Schellenberger, G., Ogorzalek, A., Oppenheimer, B. D., Stern, J. et al.\ 2023, \apj\ (subm.), arXiv:2307.01269.

\bibitem{Bogdan23}
Bogd\'{a}n, \'{A}., Khabibullin, I., Kov\'{a}cs, O. E., Schellenberger, G., ZuHone, J. et al.\ 2023, \apj, 953, 42.

\bibitem{Zhang24}
Zhang, C., Zhuravleva, I., Markevitch, M., et al.\ 2023, \mnras\ (subm.), arXiv:2310.02225.

\bibitem{Markevitch23}
Markevitch, M. et al.\ 2023, White Paper

\bibitem{Villaescusa21}
Villaescusa-Navarro, F., Angl'{e}es-Alc\'{a}zar, D., Genel, S., Spergel, D. N., Somerville, R. S. et al.\ 2021, \apj, 915, 71.

\bibitem{Werner13}
Werner, N., Urban, O., Simionescu, A., \& Allen, S. W.\ 2013, \nat, 502, 656.

\bibitem{Urban17}
Urban, O., Werner, N., Allen, S. W., Simionescu, A., \& Mantz, A.\ 2017, \mnras, 470, 4583.

\bibitem{Sarkar22}
Sarkar, A., Su, Y., Truong, N., et al.\ 2022, \mnras, 516, 3068.

\bibitem{Lodders09}
Lodders, K., Palme, H., \& Gail, H.-P.\ 2009, Landolt B\"{o}rnstein, 4B, 712.

\bibitem{Biffi17}
Biffi, V., Planelles, S., Borgani, S., Fabjan, D., Rasia, E. et al.\ 2017, \mnras, 468, 531.

\bibitem{Biffi18a}
Biffi, V., Planelles, S., Borgani, S., Rasia, E., Murante, G. et al.\ 2018, \mnras, 476, 2689.

\bibitem{Angelinelli22}
Angelinelli, M., Ettori, S., Dolag, K., Vazza, F., \& Ragagnin, A.\ 2022, \aap, 663, L6.

\bibitem{Madau14}
Madau, P., \& Dickinson, M.\ 2014, \araa, 52, 415.

\bibitem{Hickox18}
Hickox, R. C., \& Alexander, D. M.\ 2018, \araa, 56, 625.

\bibitem{Friedmann18}
Friedmann, M., \& Maoz, D.\ 2018, \mnras, 479, 3563.

\bibitem{Maoz12}
Maoz, D., Mannucci, F., \& Brandt, T. D.\ 2012, \mnras, 426, 3282.

\bibitem{Truong19}
Truong, N., Rasia, E., Biffi, V., Mernier, F., Werner, N. et al. G.\ 2019, \mnras, 484, 2896.

\bibitem{Angelinelli23}
Angelinelli, M., Ettori, S., Dolag, K., Vazza, F., \& Ragagnin, A.\ 2023, \aap, 675, A188.

\bibitem{Mernier18a}
Mernier, F., de Plaa, J., Werner, N., Kaastra, J. S., Raassen, A. J. J., et al.\ 2018, \mnras, 478, L116.

\bibitem{Eckert21}
Eckert, D., Gaspari, M., Gastaldello, F., Le Brun, A. M. C., \& O'Sullivan, E.\ 2021, Universe, 7, 142.

\bibitem{Lovisari21}
Lovisari, L., Ettori, S., Gaspari, M., \& Giles, P. A.\ 2021, Universe, 7, 139.

\bibitem{Gastaldello21}
Gastaldello, F., Simionescu, A., Mernier, F., Biffi, V., Gaspari, M. et al.\ 2021, Universe, 7, 208.

\bibitem{Nomoto13}
Nomoto, K., Kobayashi, C., \& Tominaga, N.\ 2013, \araa, 51, 457.

\bibitem{Simionescu15}
Simionescu, A., Werner, N., Urban, O., Allen, S. W., Ichinohe, Y., \& Zhuravleva, I.\ 2015, \apjl, 811, L25.

\bibitem{Mernier17}
Mernier, F., de Plaa, J., Kaastra, J. S., Zhang, Y.-Y., Akamatsu, H. et al.\ 2017, \aap, 603, A80.

\bibitem{Steidel05}
Steidel, C. C., Adelberger, K. L., Shapley, A. E., Erb, D. K., Reddy, N. A., \& Pettini, M.\ 2005, \apj, 626, 44.

\bibitem{Matsuda05}
Matsuda, Y., Yamada, T., Hayashino, T., Tamura, H., Yamauchi, R. et al.\ 2005, \apjl, 634, L125.

\bibitem{LeFevre15}
Le F\`{e}vre, O., Tasca, L. A. M., Cassata, P., Garilli, B., Le Brun, V. et al.\ 2015, \aap, 576, A79.

\bibitem{Toshikawa16}
Toshikawa, J., Kashikawa, N., Overzier, R., Malkan, M. A., Furusawa, H. et al.\ 2016, \apj, 826, 114.

\bibitem{Hatch11}
Hatch, N. A., Kurk, J. D., Pentericci, L., Venemans, B. P., Kuiper, E. et al.\ 2011, \mnras, 415, 2993.

\bibitem{Casey15}
Casey, C. M., Cooray, A., Capak, P., Fu, H., Kovac, K. et al.\ 2015, \apjl, 808, L33.

\bibitem{Overzier16}
Overzier, R. A.\ 2016, \aapr, 24, 14.

\bibitem{Chiang13}
Chiang, Y.-K., Overzier, R., \& Gebhardt, K.\ 2013, \apj, 779, 127.

\bibitem{Umehata19}
Umehata, H., Fumagalli, M., Smail, I., Matsuda, Y., Swinbank, A. M. et al.\ 2019, Science, 366, 97.

\bibitem{Wang16}
Wang, T., Elbaz, D., Daddi, E., Finoguenov, A., Liu, D. et al.\ 2016, \apj, 828, 56.

\bibitem{Tozzi22}
Tozzi, P., Gilli, R., Liu, A., Borgani, S., Lepore, M. et al.\ 2022, \aap, 667, A134.

\bibitem{DiMascolo23}
Di Mascolo, L., Saro, A., Mroczkowski, T., Borgani, S., Churazov, E. et al.\ 2023, \nat, 615, 809.

\bibitem{Vogelsberger18}
Vogelsberger, M., Marinacci, F., Torrey, P., Genel, S., Springel, V. et al.\ 2018, \mnras, 474, 2073.

\bibitem{Werner06}
Werner, N., de Plaa, J., Kaastra, J. S., Vink, J., Bleeker, J. A. M. et al.\ 2006, \aap, 449, 475.

\bibitem{dePlaa07}
de Plaa, J., Werner, N., Bleeker, J. A. M., Vink, J., Kaastra, J. S., \& M\'{e}ndez, M.\ 2007, \aap, 465, 345.

\bibitem{Mernier16b}
Mernier, F., de Plaa, J., Pinto, C., Kaastra, J. S., Kosec, P. et al.\ 2016, \aap, 595, A126.

\bibitem{deGrandi09}
de Grandi, S., \& Molendi, S.\ 2009, \aap, 508, 565.

\bibitem{Mao19}
Mao, J., de Plaa, J., Kaastra, J. S., Pinto, C., Gu, L. et al.\ 2019, \aap, 621, A9.

\bibitem{Mernier22}
Mernier, F., Werner, N., Su, Y., Pinto, C., Grossov\'{a}, R. et al.\ 2022, \mnras, 511, 3159.

\bibitem{Renzini14}
Renzini, A., \& Andreon, S.\ 2014, \mnras, 444, 3581.

\bibitem{Ghizzardi21}
Ghizzardi, S., Molendi, S., van der Burg, R., De Grandi, S., Bartalucci, I. et al.\ 2021, \aap, 652, C3.

\bibitem{Bastian10}
Bastian, N., Covey, K. R., \& Meyer, M. R.\ 2010, \araa, 48, 339.

\bibitem{Smith20}
Smith, R. J.\ 2020, \araa, 58, 577.

\bibitem{Yan21}
Yan, Z., Je\v{r}\'{a}bkov\'{a}, T., \& Kroupa, P.\ 2021, \aap, 655, A19.

\bibitem{Bulbul12}
Bulbul, G. E., Smith, R. K., Foster, A., Cottam, J., Loewenstein, M. et al.\ 2012, \apj, 747, 32.

\bibitem{Ezer17}
Ezer, C., Bulbul, E., Nihal Ercan, E., Smith, R. K., Bautz, M. W. et al.\ 2017, \apj, 836, 110.

\bibitem{McCammon2002}
McCammon, D., Almy, R., Apodaca, E., Bergmann Tiest, W., Cui, W. et al.\ 2002, \apj, 576, 188.

\bibitem{Hickox2006}
Hickox, R. C., \& Markevitch, M.\ 2007, \apjl, 661, L117.

\bibitem{LEMASS}
\lem\ All-sky SWG, et al., White Paper (in prep.)

\bibitem{Cucchetti18}
Cucchetti, E., Pointecouteau, E., Peille, P., Clerc, N., Rasia, E. et al.\ 2018, \aap, 620, A173.

\bibitem{Mernier20}
Mernier, F., Cucchetti, E., Tornatore, L., Biffi, V., Pointecouteau, E. et al.\ 2020, \aap, 642, A90.

\bibitem{Karakas10}
Karakas, A. I.\ 2010, \mnras, 403, 1413.

\bibitem{Seitenzahl13}
Seitenzahl, I. R., Ciaraldi-Schoolmann, F., R\"{o}pke, F. K., Fink, M., Hillebrandt, W. et al.\ 2013, \mnras, 429, 1156.

\bibitem{Anders89}
Anders, E., \& Grevesse, N.\ 1989, \gca, 53, 197.

\bibitem{Boylan-Kolchin09}
Boylan-Kolchin, M., Springel, V., White, S. D. M., Jenkins, A., \& Lemson, G.\ 2009, \mnras, 398, 1150.

\bibitem{Peng23}
Peng, B., Vishwas, A., Stacey, G., Nikola, T., Lamarche, C. et al.\ 2023, \apjl, 944, L36.

\bibitem{Zhang23}
Zhang, S., Cai, Z., Xu, D., Afruni, A., Wu, Y. et al.\ 2023, \apj, 952, 124.

\bibitem{Reiprich02}
Reiprich, T. H., \& B\"{o}hringer, H.\ 2002, \apj, 567, 716.



\end{thebibliography}
\end{document}